\documentclass[fleqn,usenatbib]{mnras}

\usepackage{newtxtext,newtxmath}

\usepackage[T1]{fontenc}

\DeclareRobustCommand{\VAN}[3]{#2}
\let\VANthebibliography\thebibliography
\def\thebibliography{\DeclareRobustCommand{\VAN}[3]{##3}\VANthebibliography}

\usepackage{graphicx}	
\usepackage{amsmath}	
\usepackage{tabularx}
\usepackage{orcidlink}

\newcommand{\MLV}{M_\mathrm{LV}} 
\newcommand{\K}{\mathrm{K}} 
\newcommand{\zlow}{z=0}	%
\newcommand{\zhigh}{z\simeq15}	%
\newcommand{\rvirzlow}{R_{\mathrm{vir}}^{z=0}}	
\newcommand{\mvirzlow}{M_{\mathrm{vir}}^{z=0}}	

\newcommand{\mstar}{M_{\star}}

\newcommand{\ropt}{R_{\rm opt}}

\newcommand{\tdAmin}{t_{\delta A}^{\rm min}}
\newcommand{\tdAmax}{t_{\delta A}^{\rm max}}
\newcommand{\tfmin}{t_{f}^{\rm min}}
\newcommand{\tfmax}{t_{f}^{\rm max}}

\newcommand{\tdAmaxDM}{t_{\delta A, \rm DM}^{\rm max}}
\newcommand{\tdAmaxCB}{t_{\delta A, \rm CB}^{\rm max}}

\newcommand{\tfminDM}{t_{f, \rm DM}^{\rm min}}

\newcommand{\tfmaxDM}{t_{f, \rm DM}^{\rm max}}
\newcommand{\tfmaxCB}{t_{f, \rm CB}^{\rm max}}

\newcommand{\Gyr}{{\rm \,Gyr}}

\title[Cosmic evolution around galaxy assembly sites]{Mass density structuring  around galaxy formation sites: impact on galaxy  basic properties}

\author[S. Robles, R. Dom\'inguez-Tenreiro and S. E. Pedrosa]{
Sandra Robles\,\orcidlink{0000-0002-6046-8217},$^{1,2,3,4}$\thanks{E-mail: srobles@fnal.gov}
Rosa Dom\'inguez-Tenreiro,$^{1,5}$
and Susana E. Pedrosa$^{6}$
\\
$^{1}$Departamento de F\'isica Te\'orica, Universidad Aut\'onoma de Madrid, E-28049 Cantoblanco, Madrid, Spain\\
$^{2}$Theoretical Particle Physics and Cosmology Group, Department of Physics, King’s College London, Strand, London, WC2R 2LS, UK
\\
$^{3}$Astrophysics Theory Department, Theory Division, Fermi National Accelerator Laboratory, Batavia, Illinois 60510, USA\\
$^{4}$Kavli Institute for Cosmological Physics, University of Chicago, Chicago, Illinois 60637, USA
\\
$^{5}$Centro de Investigaci\'on Avanzada en F\'isica Fundamental, Universidad Aut\'onoma de Madrid, E-28049 Cantoblanco, Madrid, Spain\\
$^{6}$Instituto de Astronom\'ia y F\'isica del Espacio, CONICET-UBA, Casilla de Correos 67, Suc. 28, 1428, Buenos Aires, Argentina
}

\date{Accepted 2026 January 19. Received 2026 January 19; in original form 2025 November 23}

\pubyear{2026}

\begin{document}
\label{firstpage}
\pagerange{\pageref{firstpage}--\pageref{lastpage}}
\maketitle



\begin{abstract}
We study  the local evolution of the Universe around galaxy formation sites in the EAGLE50 large-volume reference simulation.  
Using the reduced inertia tensor (r-TOI), we  followed the anisotropic evolution of initially  spherical Lagrangian volumes (LVs) centred at galaxy formation sites, both in dark matter (DM) and in cold baryons (CB), from very high redshift $z=15$ onward. 
We describe LV deformation in terms of the r-TOI eigen-directions, principal axes, their derived shape parameters, and the timescales for the freezing-out of these principal directions and axes. 
Of particular interest  are  the  age of the Universe, $t_{\rm U}$, when the local Cosmic Web (CW) spine emerges, and that  when  anisotropic DM mass arrangements (i.e., migrant mass flows) cease. 
We find that the shapes LVs acquire along their evolution  affect  the  halo and stellar mass of their central galaxy: 
prolate-shaped  LVs show a tendency to host low-mass galaxies at $z=0$, while massive galaxies tend to form within triaxial or oblate LVs.  Also, the local CW spine tends to set in earlier on in LVs that are to host massive galaxies than in  those harbouring less massive galaxies. 
In addition,  anisotropic DM-mass rearrangements  stop late on average, at $t_{\rm U}\sim10.5\Gyr$, and even slightly later for CB. 
Interestingly, $z=0$ LVs with either flattened configurations in CB or those that are highly prolate in DM, are  more likely to host rotation-dominated galaxies. This effect increases from $z=1$ to $z=0$. 
Finally, the CB spine of LVs that are more likely to host rotation-dominated galaxies emerges at later times.  

\end{abstract}





\section{Introduction}
\label{sec:introduction}

The  aim of this paper is to study the local evolution and structuring of the Universe around galaxy formation sites from very high redshift onward. We look for evolutionary patterns linked to the local development of the Cosmic Web (CW), patterns characterized by parameters whose statistics  and possible mutual relationships 
we  aim to disentangle.
The CW is currently observable at scales where most galaxies are bound into groups and clusters, or delineate walls and filaments  that connect them and border voids of a few tens of Mpc in diameter in position space \citep{Joeveer:1977,deLapparent:1986,Einasto:2011,pomarede:2017}.   The $z \sim 0$ CW is also observable as supercluster structures,  delineated by velocity flows, as our own  Laniakea supercluster, first discovered by \cite{Tully:2014} and more recently studied by \cite{Dupuy:2023} \cite[see also][for the first theoretical framing of $z=0$ CW observations]{Bond:1996}. However,  numerical simulations made it clear that mass elements that currently form galaxies or clusters (or their haloes) were  organized  at high redshift as a shorter-scale CW, as first shown by \citet{Klypin:1983}.

There is  increasing evidence that the CW morphological evolution    shapes  some halo and galaxy properties,  
i.e., that an interplay between the large-scale environment of a given galaxy (or halo) and some of this galaxy properties exists.
  Examples of these galaxy properties are: star formation \citep[e.g., ][]{Peng:2012,Darvish:2017,Kuutma:2017}, galaxy quenching \cite[e.g.,][]{Hasan2023,Hasan2024}, 
fractions of red  \citep[e.g., ][]{Wang-Mo:2016}  and blue galaxies \citep[e.g., ][]{LuberHasan2025}, orientation \citep[e.g., ][]{Zang-Wang-Wang:2013}, and spin direction   \citep[see e.g., ][]{Zhang-Luo:2015,Pahwa:2016}.  


 Another example of a possible role played by the CW in this respect 
 is the acquisition of angular momentum by gas, and its transport through filaments, as it travels towards halo formation regions \citep{pichon11,Codis15,Kraljic2020}.
A particular mention deserves the predictions on halo-spin alignments with the CW elements  at high redshift, 
coming from 
 the so-called Tidal Torque Theory 
\citep[TTT, see e.g.,][for a review]{Peebles:1969,Doroshkevich:1970,White:1984, Schafer:2009}, 
and its revision to include the effects of the CW anisotropic configuration \citep{Codis15,Kraljic2020}. Note however that  
some  discrepancies of the TTT predictions about low-redshift angular momentum with respect to results from numerical simulation  have recently been reported \citep{LopezVdW2025}.

Mass flows within  CW  structures also have an impact on the distribution of matter around galaxies and haloes.
For instance, some studies focused on the anisotropic character of subhalo/satellite capture along filaments \citep{
Libeskind14,Wang14,Kang15,Tempel15,Dupuy22},
as well as its possible consequences on the distribution  of satellite systems \citep{Tempel15,Wang20} and their orbital  coplanarity \citep{Benjouali11,Goerdt13,Buck15}.
On the other hand, other studies have analysed alignments between the principal directions  of the inertia ellipsoid of satellite systems  and those of shells of a scale of a  few-virial radii
 around their respective  host galaxies \citep{Shao16}, or between principal directions and the axes of slowest collapse in the matter distribution at larger scales \citep{Libeskind2015,Libeskind19}. 
An analysis of dark matter only (DMO) simulations led  to find  satellite galaxy orbital pole alignments with the intermediate  axis of the  shear field tensor\footnote{The shear tensor is the spatial rate of variation  of the deformation tensor.}~\citep{Libeskind12}. 
\citet{Welker18} studied alignments of satellite galaxies  with filaments in their neighbourhood in the Horizon-AGN simulation.

In their study of the physical processes behind the early formation of kinematic persistent planes (KPPs) of satellite galaxies,  \citet{Gamez-Marin2024} and Gámez-Marín et al. (in preparation) found alignments between the principal directions of inertia of the local density field and the orbital poles of satellites that form KPPs.
Applying the reduced inertia tensor  method~\citep{Cramer} to two zoom-in simulations,  their analysis of  the density and mass flows evolution around  galaxy formation sites   led to the conclusion that an early channel for KPP formation exists, driven by the evolution of the local density field as the local CW is established. 
An extension of this analysis to the larger-volume, higher-resolution MW/M32-like sample of galaxies \citep{Pillepich24_MWM31} taken from the Illustris TNG50 simulation  \citep{Pillepich19,Nelson19b} 
has confirmed previous findings. Furthermore, the high fraction of KPPs identified in this sample  alleviate the previous tension between simulations and MW/M31 satellite data \citep{GamezMarin2025}. 

Current insight on the  emergence and evolution of the CW stems from  analytical  studies using the Zeldovich’s Approximation \citep[][]{Zeldovich:1970} and its extension to the Adhesion Model \citep[see e.g.][and references therein]{Gurbatov:1989,Kofman92,Gurbatov:2012},
as well as via  numerical  simulations \citep[][and references therein]{Cautun:2014}.
%
These studies predict filament and wall formation\footnote{The so-called collapse to a wall or to a filament appearing in Zel'dovich's  theory plus the Adhesion Model (see references in the text) corresponds to the formation of a caustic, a mathematical singularity where no mention is made of the statistical behaviour of the particles involved.}.
As we shall see in Section~\ref{sec:GenericsCW}, it has been shown that the  morphology of the CW comes from a  hierarchical, multiscale, anisotropic collapse, where  large-scale flattened structures, frequently containing coplanar filaments  \citep[see e.g.][]{AragonCalvo2010}  are  one of its elements, 
frequently associated with  voids pushing matter outwards \citep{Kugel2024}. 
%
The driving mechanism behind the cosmic web assembly is the formation of singularities \citep{Arnold1982} at high redshift. These singularities become mass attractors, thereby guiding mass flows from lower-density regions towards them (i.e., the so-called migrant mass flows), transforming, in this way, quasi-homogenous organized matter into a CW-like configuration.

 Large-scale mass flows converging towards  galaxy formation sites are of particular interest in this work.  Thus,  the analysed scales  need to be large enough to contain high redshift proto-galaxy elements  before they are assembled into a galaxy,
as well as the mass surrounding them  along their  assembly, presumably responsible for the properties galaxies acquire along its formation and evolution.  This kind of processes  has not been fully addressed through  hydrodynamic  simulations yet, and might be a key to unveil the physical origin of some basic galaxy properties, such as mass and rotational support. 
%

 Aside from the issues found in the TTT predictions about halo-spin alignments with the CW \citep[see e.g.,][]{LopezVdW2025}, translating these predictions into galaxy-spin alignments at lower redshift demands the consideration of non-linear, baryonic physics, a complex, difficult task, that could blur TTT predictions on early halo-spin alignments. 
 In view of these difficulties, studying  galaxy-spin orientations with respect to the CW elements is not endowed with a particular meaning or interest. Thereby,  we will  focus instead on the study of galaxy rotational support.  This is quantified using the so-called kinematic morphological parameter $\kappa^{\rm rot}$  \citep{Sales:2012}, a measure of the fraction of the galaxy kinetic energy in coherent, ordered rotation, i. e., the fraction of mass moving in  close to planar and circular orbits.

In this context, apart from the above-mentioned processes of angular-momentum acquisition and transport towards galaxy formation sites, other hints about  angular-momentum causation by the CW come from \citet{Sales:2012}'s findings. They showed that the kinematic morphology of a galaxy is largely determined by the convenient  alignment of the angular momentum of baryons accreted over time,  that feed on the proto-galaxy,  with its instantaneous spin at accretion time. 
Thus, angular momentum at accretion time needs to be not only high, but also its different carriers need to be endowed with kinematically coherent angular momentum poles, so that their vectorial sum does not add up to zero or to a low total value at this time. To this end, favourable configurations would be (i) that carriers move on a flattened structure before accretion, or (ii)  that they approach the galaxy formation site spinning within an elongated structure or filament. In both situations, proto-galaxy elements would have their orbital poles clustered along conveniently long-time intervals.
As mentioned above, it turns out that the CW evolution is thought to provide flattened and elongated structures on ever larger scales at particular locations and Universe age intervals.  At these locations,  proto-galaxy elements, namely cold gas, would have their orbital poles clustered on a long-term basis. 

The specific aim of this paper is to explore  the evolution of the local density field  around galaxy formation sites, looking at the shape it acquires (e.g., is this local density field deformed into flattened and/or elongated structures?) and at its orientation, as well as the timescales for their evolution  (how  their evolution proceeds).  Next, 
we look for 
simple relations among the characteristics of the evolution of the  local density field  around a given forming galaxy and the properties that this galaxy acquires. We focus only on  basic properties, such as the galaxy mass and its  kinematic  morphological parameter, a measure of the galaxy total kinetic energy fraction in ordered rotation \citep{Sales:2012}.   

We tackle this issue by making use of a large-volume hydrodynamical cosmological simulation. 
Our methodology is
  based on previous work on the evolution of the local CW around forming galaxies \citep{Hidding:2014,Robles:2015,Gamez-Marin2024}. 
 Specifically, we use the so-called Lagrangian volumes  (LVs) constructed around these sites, and analyse their evolution by means of  the reduced inertia tensor  \citep{Cramer}.  
  In this way, we characterize the local cosmic web emergence and evolution by studying the shape a LV acquires around galaxy formation sites,  changes in its orientation, as well as the timescales to reach its final configuration.

The paper is organized as follows. In Section~\ref{sec:GenericsCW},  we outline the  basics of the CW dynamics and evolution. A summary of the method used and the large-volume simulation analysed are described in Section~\ref{sec:SimMeth}. The evolution of the local density field around forming galaxies is studied in Section~\ref{sec:LVEvol}. Section~\ref{sec:RolLVHost} is devoted to study the relationships among the LV evolution parameters, on the one hand, and  the galaxy halo and stellar masses,  and their kinematic morphological parameters, on the other hand. 
In Section \ref{sec:Discussion}, we discuss and compare our results with other findings in the literature.
In Section~\ref{sec:SumConclu}, we present a  summary of our findings and our main conclusions.



\section{The local cosmic web around forming galaxies}
\label{sec:GenericsCW}

\subsection{Theory}
\label{sec:theory}

To overcome the failure of the linear perturbation theory when the density contrast approaches unity, Zel'dovich proposed a model in which, instead of following the evolution of the density
contrast, one follows the particle trajectories \citep[][]{Zeldovich:1970}. In this way, the theory validity extends up to
the advanced non-linear stages of gravitational instability (i.e., the so-called Zel'dovich Approximation, ZA hereafter).

The ZA  is given by the {\it Lagrangian map}: 
$\vec x(\vec q,t) = \vec q + D_{+}(t) \vec{s}(\vec q)$
in co-moving coordinates, where $\vec{q}$ and $\vec{x}$ are co-moving  Lagrangian and Eulerian coordinates of fluid elements, respectively;  
$D_{+}(t)$ is the growing density mode from the linear gravitational instability theory, 
and $\vec{s}$ is the displacement field, $\vec{s}(\vec{q})=\vec{\nabla}_{\vec{q}} \phi$, with $\phi$ the displacement potential. 
In physical coordinates, the Lagrangian map, completed with mass conservation,  at the $\vec{r} = a(t) \vec{x}$ location, with $a(t)$ the expansion factor, results in the following expression for the density field:
\begin{equation}
\rho(\vec{r},t) = \frac{\rho_b(t)}{ \prod_{i=1}^{3}[1-D_{+}(t)\alpha_i(\vec{q})]},
\label{eq:DenZAppro}
\end{equation}
where $\rho_b(t)$ is the background  density at Universe age $t$  and 
$\alpha_1(\vec{q}) > \alpha_2(\vec{q}) >  \alpha_3(\vec{q})$ are the eigenvalues of the local deformation matrix 
$\partial^2\phi/\partial q_i\partial q_j=\partial s_j/\partial q_i$, whose trace verifies 
\begin{equation}
\label{eq:landscape_height}
\vec{\nabla}_{q}\cdot\vec{s} =  \sum_{i}^{}\alpha_{i}(\vec{q}) = \frac{5}{3} \frac{\delta\rho(t_{\rm ini})}{\rho_b(t_{\rm ini})},  
\end{equation}
where $t_{\rm ini}$ is the age of the Universe  when initial conditions are taken for the evolution.  

Caustic  (i.e., a region of infinite density in Eulerian space) formation is described by Eq.~\ref{eq:DenZAppro} in the ZA.
Here, a positive (negative) eigenvalue means collapse (expansion) along the 
direction of the corresponding eigenvector.
Accordingly, the first patch of matter collapses when $D_{+}(t)\alpha_1(\vec{q}) = 1$, along the direction of the eigenvector  associated to the largest eigenvalue, resulting in a wall-like singularity or pancake. 
Next, the patch will  collapse along the  eigen-direction corresponding to the second largest eigen-value  ($D_{+}(t)\alpha_2(\vec{q}) = 1$), perpendicular to the first, giving rise to a filament,  and lastly 
along the remaining direction ($D_{+}(t)\alpha_3(\vec{q}) = 1$). This is assuming all three eigenvalues are positive.
Thus, in three dimensional (3D) space caustics can be walls, filaments or nodes.

An issue with the ZA is that, contrary to its predictions, cosmological simulations of large-scale structure formation indicate that long-lasting pancakes are indeed formed (and not blurred by multi-streaming soon after their formation), with particles sticking and multi-streaming not taking place.  
This problem with the ZA was rectified by \cite{Gurbatov:1989} and \cite{Kofman:1990} who formulated the so-called Adhesion Model. By adding a viscosity term to the ZA's momentum equation that is effectively zero anywhere but at the caustics (i.e., regions with large velocity variations), 
this model is capable of reproducing the structuring observed in numerical studies \citep{Cautun:2014}. This is,  predicting when and where caustics form. 
A review of physically motivated derivations of the `Adhesion Model' was presented by  \citet{Buchert:2005}.

Unlike in the ZA, in the AM, matter adheres to each other giving rise to dense regions in Eulerian space, instead of forming multi-streaming regions. 
Walls are the first structures to emerge, prompting mass flows towards them. Next,  filaments are formed within walls attracting anisotropic mass flows. Then, filaments become the paths these flows follow towards nodes,  where matter finally piles up \citep[i.e., the Cosmic Web elements, see][]{Bond:1996}.

   It is worth noting that Eq.~\ref{eq:landscape_height} entails that the eigenvalues $\alpha_i(\vec{q})$ in a local patch are related to the intensity of the initial fluctuation field on that patch. As a result,  the process of structure formation described above will be delayed in patches whose fluctuation field is low \citep{Hidding:2014,Robles:2015}.
%
We will find such situations along the paper, in particular to explain the ordering of freezing-out timescales for different patches of matter.

\subsection{Numerical simulations and observations: the baryonic cosmic web}
\label{sec:simusobserv}

The ZA and AM do neither predict/describe caustic inner properties nor gas hydrodynamic and thermodynamic behaviour. 
Numerical simulations are needed to this end. In a cosmological context, dark matter (DM) only simulations have been the first methodology to be used, e.g., the Millenium \citep{Springel2005}, the Multidark and Bolshoi simulations \citep{Klypin-Multidark:2016,RodriguezPuebla}, see \cite{Angulo2022} for a review.

DMO simulations were the first to show that piled mass relaxes and reaches equilibrium  forming haloes. The other CW elements also collapse once the web-like mass configuration,  at the considered scale, turns around 
\citep[see][]{Bond:1996}\footnote{It is worth noting that when collapse to a CW filament or a wall is mentioned along this paper, this implies in 
 general not a simple, mathematical   caustic formation, but  a multi-scale, complex, non-relaxed structure, made of  different smaller-scale CW elements. }. This erases the CW structure below this scale, scale whose size at a given  age of the Universe  depends on the location.
At the same time, this web organization, with  walls and filaments  still developing,  emerges at larger and larger scales. Indeed, coherent large-scale cosmic flows of galaxies within attractor basins have very recently been observationally detected at supercluster scales and beyond \citep[see e.g.][and references therein]{ValadeAttracBasins:2024,Hoffman-CosFlows4:2024}.
As just mentioned,  we empirically know that, at a given scale, CW elements are morphologically transient structures.

The details of the CW building up by gravitational forces, including how the shaping of the CW occurs, are  not known in depth yet.  
Details on the so-called `baryonic CW',  framed by the DM CW,  are  even more poorly known. This CW is basically made of gas, and it includes,  apart from the interstellar medium (ISM, gas in galaxies), the so-called circumgalactic medium (CGM), for a review see \citet{Tumlinson:2017}, and the  intergalactic medium (IGM), i.e., gas beyond galaxies and  within and outside haloes, respectively. See \citet{McQuinn2016} for a review of the IGM, and more recently, e.g., \cite{Tejos:2026}, and references therein\footnote{
The IGM comprises two phases: (i) the warm-hot intergalactic medium   (WHIM; see e.g., \citet{DaveCenOstriker:2001}) a mild-density gas distribution shock-heated to a temperature of $10^5 K < T < 10^7 K$, and  (ii), the often dubbed diffuse IGM,  gas photo-ionized and photo-heated to a temperature of $T < 10^5 K$. While the WHIM dominates the baryon budget at $z<0.5$, the diffuse IGM does it at $z>2$. \label{foot:gas}}.

As for the CGM,  a disc of rotating cold gas, close to and well aligned with the stellar disc plane, with high tangential velocities 
has been identified in different  hydrodynamical simulations   
\citep[see e.g.,][]{Danovich:2012,Danovich:2015,Oppenheimer:2018,Ho:2019,Grand:2019,DeFelippis:2020}. 
CGM disc properties have been found to be  largely independent from the simulation considered \citep[e.g.][]{Stewart:2017}.
 Radial inflows of cold gas in coplanar inspiraling streams, that fed on the CGM rotating structure have also been found.
 
Several studies have found that the gas space configuration  does not follow the dark matter CW. Indeed, in agreement with \citet{CenOstriker:1999} and \citet{DaveCenOstriker:2001} seminal work, more recent analyses of the TNG300 simulation \citep{Nelson19a} have concluded that the large-scale gas distribution depends on the particular gas component, with the WHIM (see footnote~\ref{foot:gas})  mostly residing in large, dense filaments at $z=0$,  while the IGM mostly being in walls, see \citet{Martizzi2019,Artale2022,Walker2024}.

The study of observational aspects of the baryonic CW is still an incipient field. Using fast radio burst (FRB) detection data\footnote{FRBs are luminous signals with a duration from microseconds to milliseconds, known to traverse extragalactic distances.},
\cite{Connor2025} have recently reported  a high baryon fraction in the IGM and a dearth of baryons in the CGM, i.e., within galaxy haloes. 
These results are in agreement with the above mentioned results from simulations, as well as with kinematic
Sunyaev-Zel’dovich (kSZ) results \citep[e.g.,][]{Hadzhiyska2024}. 
Studies of the MUSE Hubble Ultra Deep Field Survey using  Lyman-$\alpha$ emission observations of the IGM at $z\sim 4 -3 $  by \cite{Tornotti2025,Tornotti2025b} pave a new avenue for constraining the physics of the baryonic CW at large scales.

These results from theory, simulations and observations,  
are interesting and useful for our work. 
Motivated by them, 
we will follow the evolution of the dark matter and the baryonic CWs  separately.


\section{Simulations and  Method}
\label{sec:SimMeth}

\subsection{Simulation}

We selected a sample of galaxies from the Evolution and Assembly of Galaxies and their Environments (EAGLE) simulation suite~\citep{Crain:2015,Schaye:2015}, namely the reference simulation with comoving cubic box length of 50  Mpc side  and $752^3$ DM particles. The number of baryonic particles is initially equal to that of the DM component.  This simulation assumes a $\Lambda$CDM cosmology with parameters from the \citet{Planck:2014}. The simulation was run using the N-body smoothed particle hydrodynamics (SPH) Lagrangian code \texttt{GADGET-3}~\citep{Springel:2005,Springel:2008}. Numerically unresolved baryonic processes are implemented using subgrid physics  modules. These include (i) radiative cooling and photoheating  \citep{Wiersma:2009a}, (ii) star formation as in \citet{Schaye:2008}, (iii) stellar mass loss and supernovae (SNe) following~\citet{Wiersma:2009b} (iv) stochastic stellar feedback \citep{DallaVecchia:2012}, and (v) black holes and feedback from AGNs~\citep{Rosas-Guevara:2015,Schaye:2015}. Free parameters of the subgrid  modules that deal with feedback from star formation and AGNs in the EAGLE reference simulations were calibrated to ensure agreement with observational data at $z\sim0$, namely the galaxy stellar mass function, galaxy sizes and  the black hole mass–stellar mass relation~\citep{Crain:2015,Schaye:2015}.

The halo catalogue was constructed by running first the friends-of-friends (FoF) algorithm~\citep{Davis:1985} on the DM particles. Baryonic particles are assigned to the same halo as their nearest DM particle, if any. Next, self-bound substructures within  FoF haloes are identified using the \texttt{SUBFIND} algorithm \citep{Springel:2001,Dolag:2009}, considering all particle species. To connect these substructures across time, merger trees were constructed using the \texttt{D-TREES} algorithm \citep{Jiang:2014,Qu:2017}. 

\subsection{Method}
\label{sec:maths}

\begin{figure}
\centering
\includegraphics[width=\columnwidth]{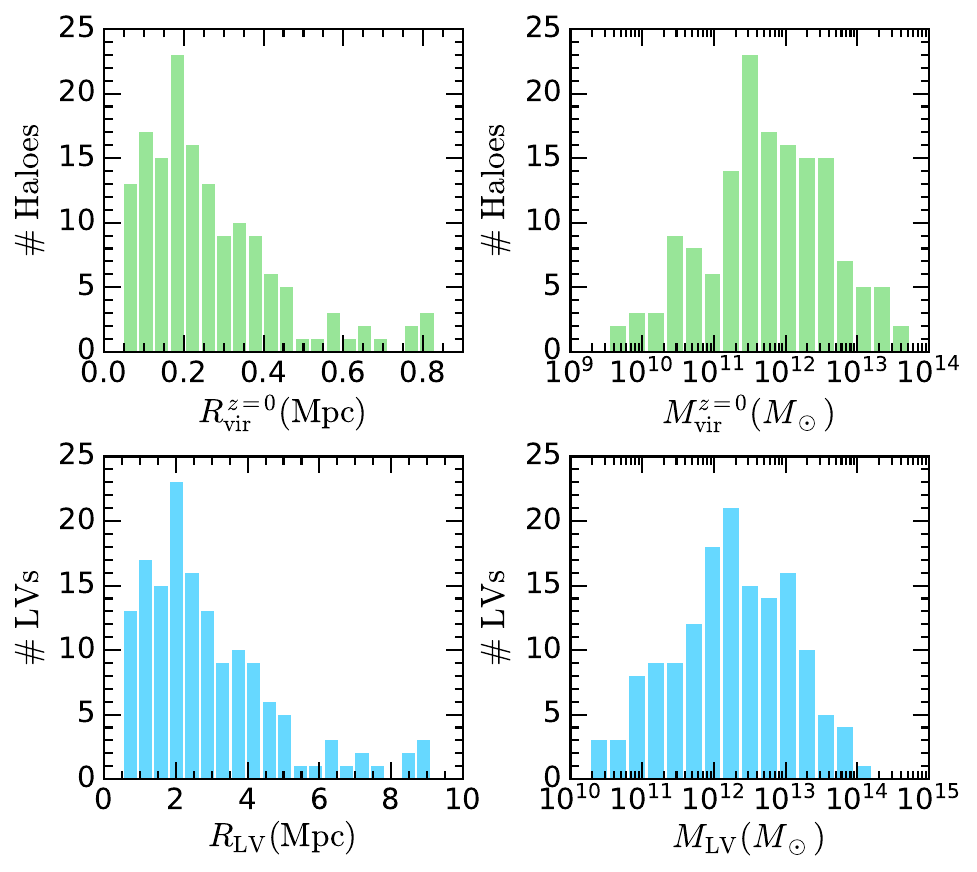} 
  \caption{Upper panels show the radius and mass distribution of the galaxy haloes selected at $z=0$. Lower panels depict 
  the mass and radius of the corresponding LV sample.}
 \label{fig:histmassrad} 
\end{figure}

To construct the LV sample, we first selected those  central galaxies at $z=0$ with an identified progenitor at $z\simeq15$. The virial radius $\rvirzlow$ and mass $\mvirzlow$ distributions of their corresponding halos are shown in the upper panels of Fig.~\ref{fig:histmassrad}. 
The virial radius is defined following the prescription given by \citet{Bryan:1998}. 
Next, we selected all particles enclosed within $K\times\rvirzlow$, with $K=10$, around each galaxy seed at $\zhigh$. The centre of every galaxy progenitor at $\zhigh$ was obtained from the EAGLE public database~\citep{McAlpine:2016}. 
We chose $K=10$ to compare our results with those of \citet{Robles:2015}, whose approach we have  followed,  and that found no relevant differences in  their findings when increasing $K$ from 10 to 15. On the other hand, a lower value of $K$ may not capture the complete process of wall formation~\citep{Robles:2015}. 

The distributions of the mass ($\MLV$) and radius ($R_\text{LV}$) of the LV sample are given in the lower panels of Fig.~\ref{fig:histmassrad}.  
We see that the $\mvirzlow$  and $M_{\rm LV}$ distributions are very similar. Indeed, a fit to their values gives $\mvirzlow = 0.23 (M_{\rm LV})^{1.01 }$, with an extremely low dispersion. This is a consequence of the protocol used to build up the LV set. It is worth remarking that, while $\mvirzlow$ has the usual physical meaning, $M_{\rm LV}$ does not possess any and as such it cannot be considered as a parameter determining LV properties. Thus, $\mvirzlow$ is a proxy for $M_{\rm LV}$, and we shall see that it depends on the shape that the LV takes along its evolution.

Once the initially spherical LVs were determined, we followed the evolution of each of them across redshift from $\zhigh$ to $\zlow$, i.e. we tracked  all DM, gas and stellar particles within each LV. These particles or mass elements sample the deformations (stretching, folding, collapse) of the local cosmic web~\citep{Robles:2015}. 
Note that by definition the LV mass is preserved in time. The sole exception being black hole (BH) formation, since we do not trace those particles.  

To quantify the above-mentioned LV deformations across redshift, we computed the reduced inertia tensor (r-TOI) of each LV relative to its centre of mass
\begin{equation}
I_{ij}^{\rm r} =\sum_{n=1}^{N}m_n\frac{(\delta_{ij}r_{n}^2 - r_{i,n}r_{j,n})}{r_{n}^2}, 
\label{eq:redineten}
\end{equation}
where $r_{n}$ is the distance of the $n$th LV particle to the LV centre of mass and $N$ is the total number of such particles. The reduced inertia tensor is invariant under mass rearrangements along the radial direction, and hence it is less affected by the presence of substructure at the outskirts of the LV~\citep{Gerhard:1983,Bailin:2005}.  
This is of particular interest to deal with anisotropic mass configurations as is our purpose here. Note that  the different gas components cannot been distinguished in detail by the r-TOI formalism.
The definition given in Eq.~\ref{eq:redineten} considers all the particles within the LV, regardless of their type. When only a subset of particle types is  considered, the same definition applies, but the centre of mass is calculated only with these particles and the distances are calculated respect to it, see terminology in Table~\ref{tab:Glossary}. 

The next step is to calculate the eigenvalues ($\lambda_1\leq\lambda_2\leq\lambda_3$) and eigenvectors ($\vec{e}_i$, $i=1,2,3$) of the reduced inertia tensor, where $\vec{e}_1$ and $\vec{e}_3$ are the major and minor axis of the inertial ellipsoid, respectively. Its principal axes ($a\geq b\geq c$) are given by
\begin{eqnarray}
a &=& \sqrt{\frac{\lambda_2 - \lambda_1 + \lambda_3}{2\MLV}}, \qquad
b = \sqrt{\frac{\lambda_3 - \lambda_2 + \lambda_1}{2\MLV}}, \\ \nonumber
c &=& \sqrt{\frac{\lambda_1 - \lambda_3 + \lambda_2}{2\MLV}}. 
\end{eqnarray}
Note that $\lambda_1 + \lambda_2 + \lambda_3 = 2\MLV$ and $a^2 + b^2 + c^2 =1$.

We have computed the principal axes and directions of the inertia ellipsoid using all the particles in each LV (tot: DM, gas and stars), and using only one of the following components: DM,  cold baryons (CB hereafter, made of cold gas and stellar particles), and hot gas. 
We have considered as hot gas particles those heated to a temperature higher than $3\times10^4\K$. Our analysis in the next sections  will focus on LV properties when considering either its DM (DM-LV) or CB (CB-LV) particle content.

Shape deformations are measured by computing the $b/a$ and $c/a$ axis-ratios, as well as the shape parameters:  triaxiality ($T$)~\citep{Franx:1991}, ellipticity ($e$) and prolateness ($p$)~\citep{Bardeen:1986,Porciani:2002,Springel:2004} of each LV across redshift,
\begin{eqnarray}
T &=& \frac{a^2-b^2} {a^2-c^2}, \label{eq:Tdef} \\
e &=& \frac{a^2-c^2}{a^2+b^2+c^2} , \qquad
p=\frac{a^2+c^2-2b^2}{a^2+b^2+c^2}.
\label{eq:ShapePDef}
\end{eqnarray}
Note that a system with $c/a>0.9$ has an almost spheroidal shape, objects with $c/a$ below this threshold and $T<0.3$ are oblate spheroids, while those with $T>0.9$  have an extremely prolate triaxial shape ($c,b\ll a$)~\citep{GonzalesGarcia:2009}. In extremely oblate systems $b/a\rightarrow1$ and in flattened elliptical structures $c/a\rightarrow0$ (i.e. $c \ll a$, with any $b/a$ value, this is any triaxiality). 
Thus, three distinguished $T$ regions can be defined: $0.7 \le T \le 1$, $0.3 \le T \le 0.7$, and $T \le 0.3 $, corresponding to  prolate, triaxial and oblate shapes (see Fig.~\ref{fig:LVevol-plane}).

Similarly, using the $e$ and $p$ parameters to characterize a system shape, nearly spherical objects have $e<0.2$ and $|p|<0.2$, extremely thin filamentary structures are those with $e\simeq p\simeq1$. Prolate systems have $p>0$, while those with $p>0.5$ can be considered as extremely prolate structures. Note that in the latter case $p>0.5$ implies that $e>0.5$, $b<1/\sqrt{6}$ and $a>\sqrt{2/3}$,  which in turn results in  $b/a<0.5$.

Whenever we consider only a subset of the particle types within the r-TOI calculation, the same definitions apply and the same relations hold, but using $M_{\rm comp}$ instead of $M_{\rm LV}$, where `comp' refers to the particle type(s) included. When the specification of the type of particles involved in the r-TOI calculation is necessary, a subindex will be added to the symbol of the principal direction or axis, or to the shape parameter symbol, see the list of terms in Table~\ref{tab:Glossary}.

Contrary to methods to analyse and classify the CW \citep[see e.g.,][and references therein]{Hoffman2012,Cautun:2013,Libeskind18}, which are local and rely on tensorial tools defined at specific points and as such require smoothing procedures, our approach is simpler and follows the evolution of the local environment where galaxies are to be formed. We track the average large-scale deformations of initially spherical LVs across the different snapshots provided by the simulation. 
To describe the LV evolution, only a few parameters are required:  three  principal directions (one of which stops significantly evolving at very early times), and the principal axes $a(t), b(t), c(t)$,  of which only two are independent. This method accurately captures the evolution of the local environment around galaxy formation sites,  and it is highly suited for our purposes here.


\section{LV evolution around forming galaxies}
\label{sec:LVEvol}

\subsection{Evolution of the principal directions}
\label{sec:PriDirecEvol}

\begin{figure}
\centering
\includegraphics[width=0.9\columnwidth]{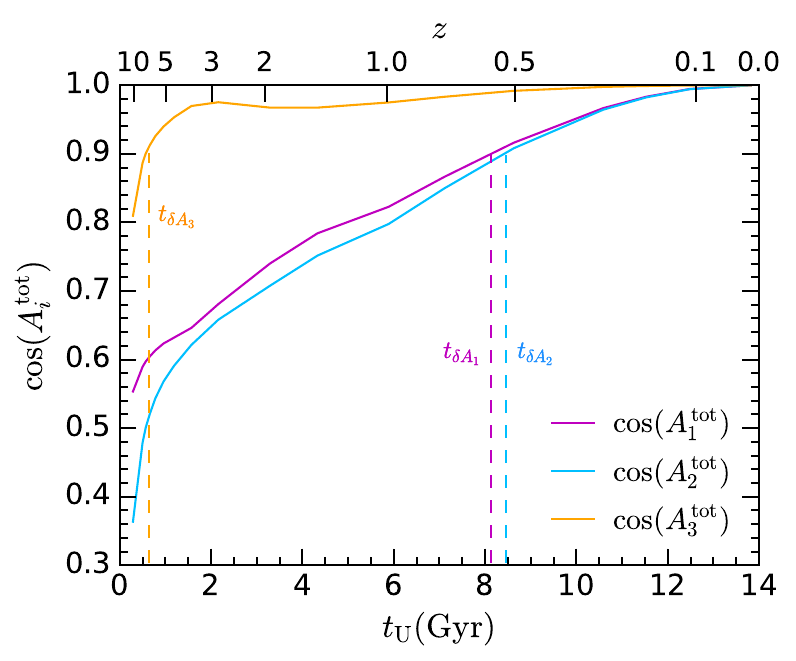} 
\caption{Example of the evolution of the three  eigen-directions of the reduced inertia tensor calculated considered all particle species (tot).}
\label{fig:angAievol}
\end{figure}

The eigen-directions of  the LV reduced inertia tensor change with cosmic time as a consequence of anisotropic mass reconfigurations. Figure~\ref{fig:angAievol} illustrates  such evolution by showing $A_i^\mathrm{tot}$ as a function of the age of the Universe $t_{\rm U}$. This is the angle formed by the eigenvectors
 $ \hat{e}_i^{\rm tot}(z)$ and $\hat{e}_i^{\rm tot}(\zlow)$, with $i=1,2,3$, and where `$\rm tot$' stands for the eigenvectors of
 the $I_{ij}^{\rm r}$, calculated with all the particles within a LV. Note that since we measure the variation in the direction, this angle varies between $0^\circ$ and $90^\circ$.

 We see that while at high redshift these angles are considerably varying, at low $z$ changes are severely dimmed and until the eigen-directions become finally frozen (within a threshold). Vertical lines mark the freezing out timescale within a threshold of $10\%$ of  $\cos(A_i^\mathrm{tot})$, see Section~\ref{sec:freezing-times} for further details.
 We also note that, in this particular case, one principal direction (i.e., $ \hat{e}_3^{\rm tot}(z)$) becomes frozen first, while the other two stop changing long after.  In the interval in-between, the LV mass rotations are restricted to just one degree of freedom.

\begin{figure}
\centering
\includegraphics[width=\columnwidth]{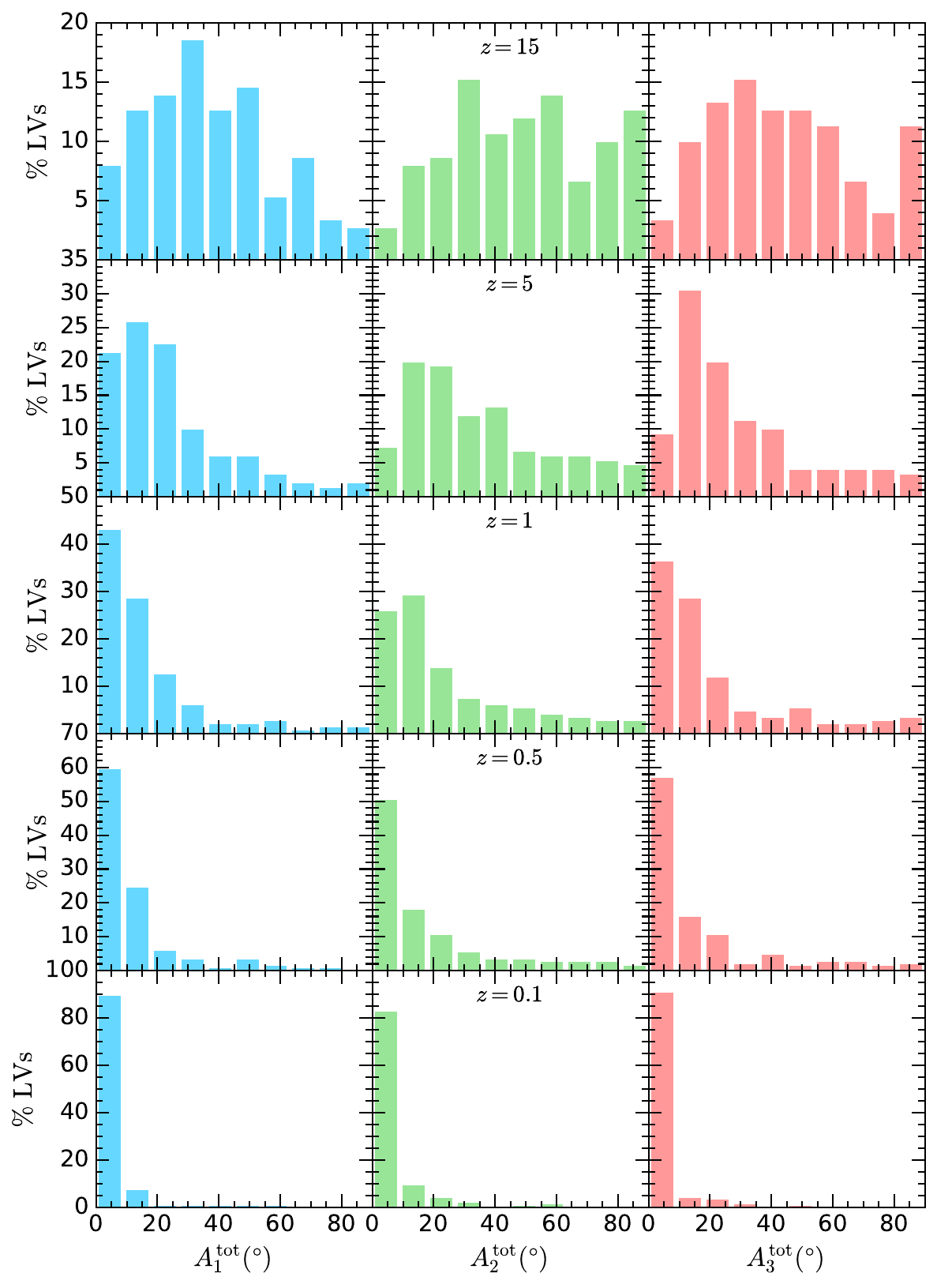}
\caption{Evolution of the distribution of eigen-directions across redshifts, where $A_i^\mathrm{tot}$ is the angle formed by the eigenvectors
 $ \hat{e}_i^{\rm tot}(z)$ and $\hat{e}_i^{\rm tot}(\zlow)$, with $i=1,2,3$, and where `$\rm tot$' stands for the eigenvectors of
 the $I_{ij}^{\rm r}$, calculated with all the particles within a LV. }
\label{fig:angAievol-Glo}
\end{figure}

 It is interesting to analyse the global  behaviour of the LVs in this respect and the extent to which its different components (DM, cold baryons and hot gas) affect this behaviour. 
In Fig.~\ref{fig:angAievol-Glo}, we show 
the  evolution of the 3 principal directions for all the LVs in our sample, calculated with  the total LV mass along cosmic time. No clear statistical preference for a particular principal direction to evolve faster than the other two shows up in the histograms. The evolution is not very fast, as at $z=0.5$, $\sim 45\%$ of the LVs have not completed their evolution yet.
 
\begin{figure}
\centering
\includegraphics[width=\columnwidth]{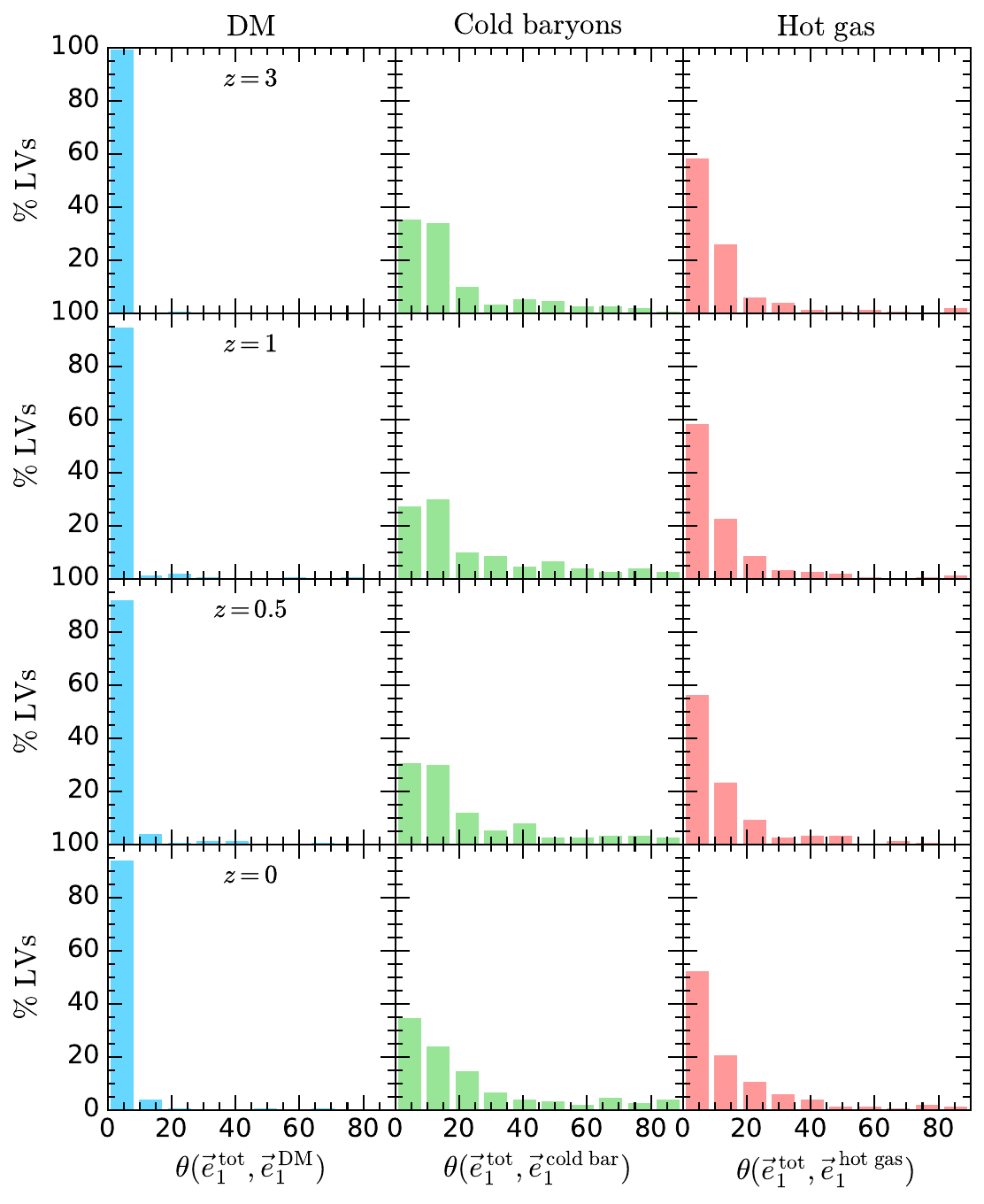} 
  \caption{Evolution of the distributions of the angles formed by the direction of the major
 axis of inertia that arises from the overall
 mass distribution  $ \hat{e}_1^\mathrm{tot}(z)$ and the corresponding eigen-direction calculated with one the following components: DM (left), cold baryons (middle), and hot gas (right).}
\label{fig:angAievol-Comp}
\end{figure}

In Fig.~\ref{fig:angAievol-Comp}, we compare the evolution of the principal directions of each of the LV components (DM, cold baryons and hot gas) with respect to that of all the particles within a LV. 
As expected, the  LV evolution is driven by the DM component. Note here  that neither the CB nor the hot gas component follow the DM evolution concerning  their respective eigen-directions. 
The behaviour of these two components is not unexpected. Indeed, the temperature of a gas particle increases due to   gravitational or shock heating, as well as via energy injection by discrete sources, e.g., supernova  or black hole feedback.
These  processes cause, among other effects, outflows from galaxies towards the CGM or even to the IGM, see e.g. \citet{Tinsley:1980-2022,King:2015ARAA,Christensen:2018,Maiolino:2019ARAA,RocaFabrega:2019,Ramesh2023}, more likely from low-mass galaxies, see e.g. \citet{DekelSilk:1986}, and  \citet{Miranda:2025}.
On the other hand, gas cools through dissipation and other processes. Both gas heating and cooling processes do not affect directly the DM component but remove or add cold gas, changing also their location,  and remove stellar particles. 
These new or removed particles are examples of  changes in the type of the particles within  a LV, that affect its CB component behaviour. 
A second effect on gas particles are the pressure from hydrodynamical forces, or  from  energy injection by discrete energy sources. Thus, shape or principal direction changes in the CB component happen not only due to the  CW setting in and its dynamical evolution, but also due to gas heating or cooling and pressure or star-forming effects as well.

\subsection{Shape Evolution}
\label{sec:LV-shape-evol}

\subsubsection{Individual LV shape evolution}
\label{sec:IndLVShape}

\begin{figure}
\centering
\includegraphics[width=\columnwidth]{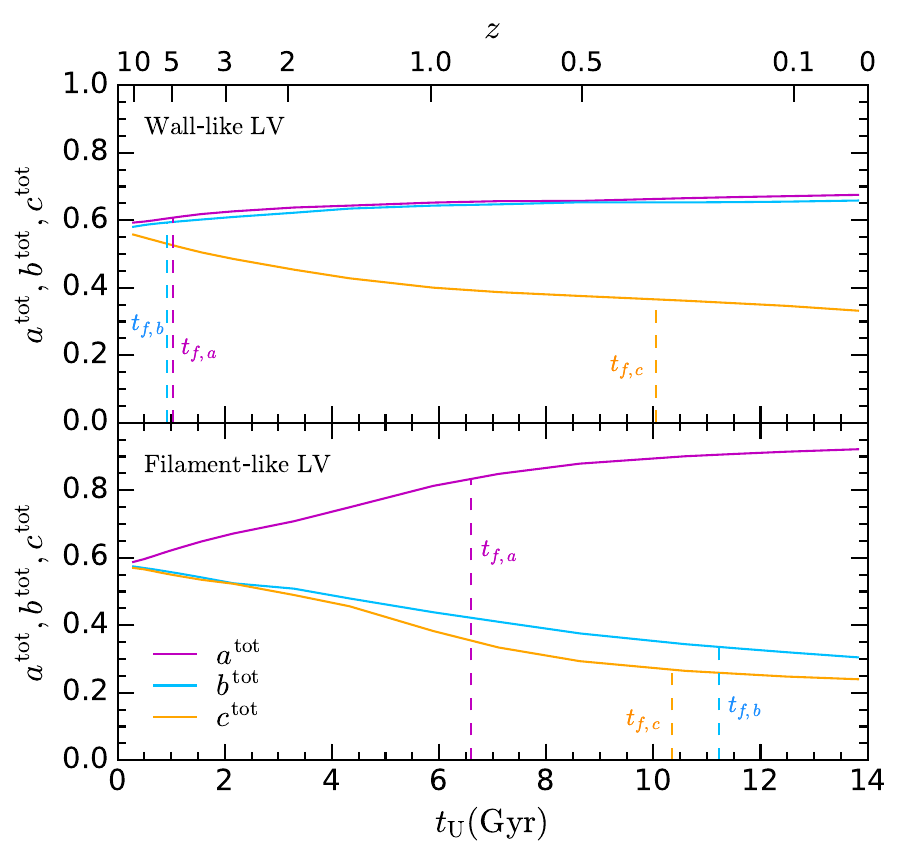} 
\caption{Evolution of the principal axes of the inertia ellipsoid for a wall-like LV (top) and a LV with evolves into a  filamentary shape (bottom), where `tot' stands for the total mass of the LV. i.e. all particles species.}
\label{fig:prinaxesevol}
\end{figure}

 The pattern of the r-TOI principal axis evolution is as follows: while the $a(t)$ axis monotonously grows and $c(t)$ shows a shrinking behaviour, the $b(t)$ principal axis can grow (deforming the LV into an oblate shaped body), decrease (giving rise to a prolate or even a needle-like configuration), or remain almost constant (triaxial shaped LVs). This last possibility occurs for $\sim$ half the LV deformation histories,  leading to an early freezing out of the $b$-axis.  This means that, in this case, the LV cannot  evolve to a prolate shape  (remember that $a(t)^2 + b(t)^2 + c(t)^2 = 1$).

Figure~\ref{fig:prinaxesevol} follows the evolution of the principal axes, $a, b$, and $c$, across time for two LVs (considering all particle types). These examples of shape evolution illustrate two possible final mass configurations of initially spherical LVs. 
The upper panel 
corresponds to a shape transformation from spherical at high redshift $z$ to an oblate (i.e., flattened)  structure at lower $z$. The lower panel shows the  evolution of the   principal axes from spherical at high $z$ to a filament-like LV at lower redshift. 
Note that, in both cases, the shape transformation proceeds faster at high redshift, and later on the deformation rate  diminishes. 
Vertical lines denote the freezing-out timescales, i.e, the age of the  Universe  when the values of each of the axes reaches its final value at $z=0$, within a 10\% (see Section~\ref{sec:freezing-times} for a precise definition).
In the case of the wall-like final configuration (upper panel), the $b$ axis freezes out first at very high redshift, i.e., within the first Gyr of the Universe evolution.  Very soon after, the $a$-axis freezes out as well.

 Some considerations are in effect to underline the appropriateness of analysing the shape evolution of local LVs  around forming galaxies.
DM-LVs are shaped by gravitational forces acting on their particles, see Section~\ref{sec:GenericsCW}. 
The CB-LVs suffer, as well, the effects of hydrodynamic forces and of energy injection from discrete sources, as mentioned before. The shape a LV acquires at a given $z$ tracks down the global effect of all these processes until this particular $z$. This is the first reason to follow LV shape evolution. 

The second reason is as follows. The geometry of mass density around a forming galaxy sets up conditions for mass accretion onto it.
For instance,  central  galaxies forming at the expense of a prolate-like shaped DM-LV or CB-LV (where $c,b\ll a$), would  have a tighter limitation in their merger and smooth mass accretion history, as their mass supply comes from a  more anisotropic structure than in the flattened and triaxial  LV cases. 
In other words, mass accretion towards  proto-galaxies in prolate LVs stem from only one direction, while mass fluxes towards those in oblate LVs can come  from  three or more filaments. This could affect the final mass of the central halo and its corresponding galaxy.

On the other hand, the LV shape could affect rotation parameters of its central galaxy as well. Indeed, a low $(c/a)_{\rm CB}$ configuration  around a forming galaxy would likely imply a higher kinematic coherence of the particles to be accreted during the formation process, and thus a higher probability of becoming discy as a result, see Sections~\ref{sec:introduction} and \ref{sec:ShapeOnKrot}\footnote{A flat shape would occur if the LV is part of a global compression of mass caused by larger scale mass configurations,
such as void expansion pushing their inner mass towards their boundaries, and even more when two voids push from opposite directions, among others \citep[see e.g.][]{Kugel2024}.}.  
In the same line,  not only flattened, but also extremely prolate configurations could play a role in  central galaxy rotational support. Indeed, as 
mentioned in Section~\ref{sec:introduction} (and references therein),  filaments are the richest angular-momentum configurations in the CW, as well as the highways to transport angular-momentum  towards galaxy formation sites, playing a particular role in the formation of galaxy discs.


\subsubsection{Statistics of the LV shape evolution}
\label{sec:ShapeEvolStat}

\begin{figure}
\centering
\includegraphics[width=\columnwidth]{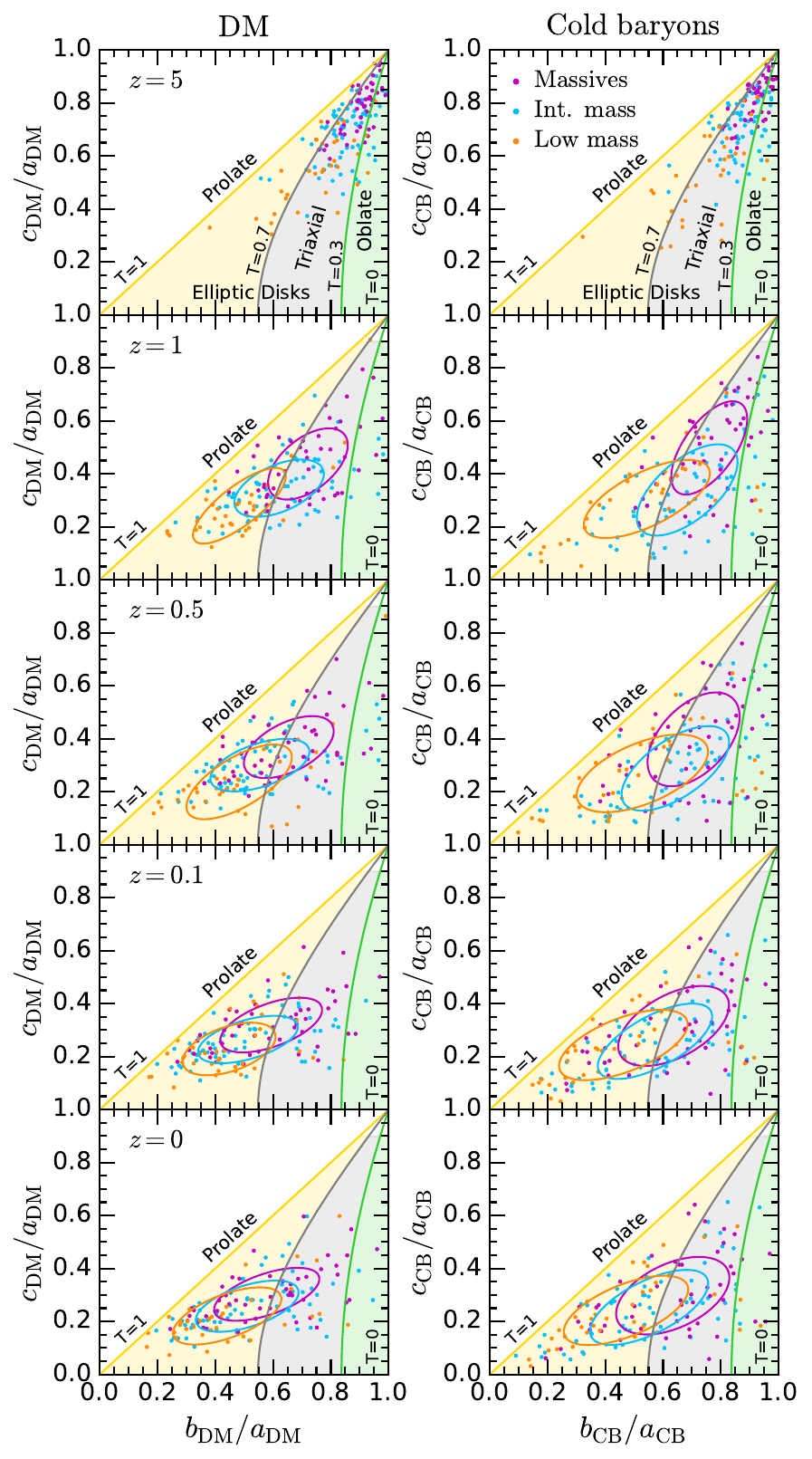}
  \caption{Shape evolution of  the DM- and the CB-LVs in the axis-ratio plane (left and right columns, respectively) across the  redshifts, $z$,  shown  in the left panels. 
  The LV sample is split in three  mass groups: 
 massive LVs with 
   $\MLV\geq5\times10^{12} M_\odot$ (magenta), LVs with intermediate mass, $5\times10^{11}\leq \MLV<5\times10^{12} M_\odot$ 
   (light blue), and low-mass LVs, $\MLV<5\times10^{11} M_\odot$ (orange). These bins correspond to the following groups in the $z=0$ central halo mass formed in each LV: $M_{\rm vir}^{z=0}\geq 1.52\times10^{12}M_\odot$, $1.48\times10^{11}M_\odot\leq M_{\rm vir}^{z=0}< 1.52\times10^{12}M_\odot$,  and $M_{\rm vir}^{z=0}<1.48\times10^{11}M_\odot$. The  region shaded in yellow denotes LVs with $0.7\leq T \leq 1$ and $c/a<0.9$. Triaxial LVs lie in the grey shaded region, while the region corresponding to oblate ellipsoids  $c/a<0.9$ and $0\leq T\leq0.3$ is shaded in green. Concentration ellipses at 1$\sigma$ for each mass group are also shown. } 
\label{fig:LVevol-plane}
\end{figure}


\begin{figure*}
\centering

\includegraphics[width=\columnwidth]{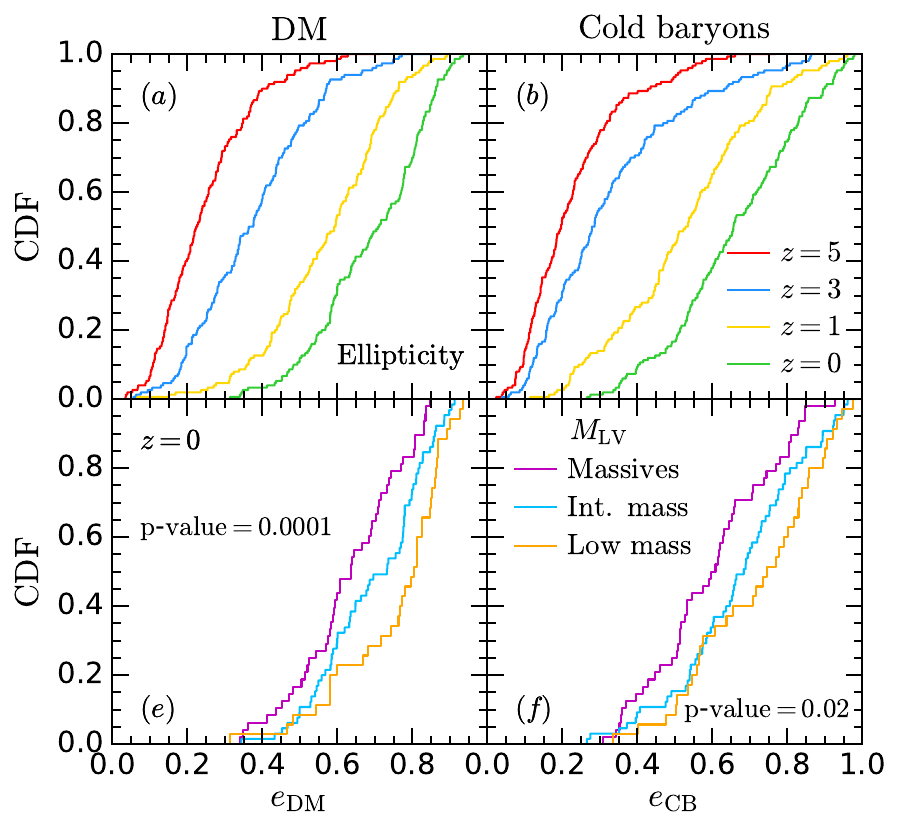}
\includegraphics[width=\columnwidth]{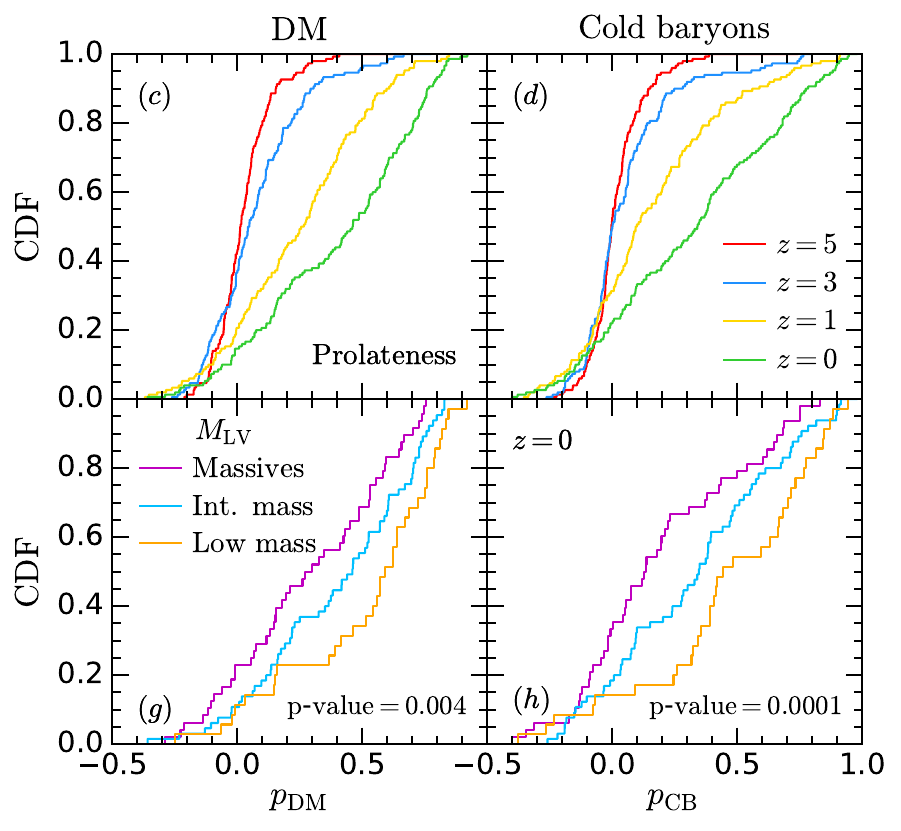}
\caption{ First row: CDF evolution for the 
ellipticity  $e$ parameter considering the DM and CB reduced inertia tensors,  panels (a) and (b) respectively. Panels (c) and (d) correspond to the evolution of the prolateness $p$ parameter. Different  colours for the CDFs stand for different redshifs. 
Second row addresses LV shape and mass relation as measured by the $e$ (panels (e) and (f)) and $p$ (panels (g) and (h)) parameters. The LV sample is binned in three groups: massive LVs  ($\MLV\geq5\times10^{12} M_\odot$, magenta), LVs with an intermediate mass ($5\times10^{11}\leq \MLV<5\times10^{12} M_\odot$, light blue) and low-mass LVs  ($\MLV<5\times10^{11} M_\odot$, orange).
The  p-values of the two-sample Kolmogorov-Smirnov (KS) test correspond to the difference between the massive and low-mass LVs. }

\label{fig:LVevol-CDF-ep}
\end{figure*}

The evolution of the LV set in the axis-ratio  $c/a$ versus $b/a$ plane, for both DM and CB LV components, is given in Fig.~\ref{fig:LVevol-plane}, where the regions corresponding to different shapes according to the values of the triaxiality  $T$ parameter  are shaded in different colours, and each point stands for a LV. Additionally, to gain insight into the possible  effect of the LV shape  on the $z=0$ halo mass, $M_{\rm vir}^{z=0}$, of the central galaxy formed within each LV,  low, intermediate mass and massive LVs are depicted in orange, light blue and magenta, respectively\footnote{We remind the reader  the tight correlation between $M_{\rm vir}^{z=0}$ and $M_{LV}$. Thus, we can consider that in Fig.~\ref{fig:LVevol-plane}   the LV set is binned  in both masses, $M_{\rm vir}^{z=0}$ and $\MLV$.}.
The same colours have been used to draw the 1$\sigma$ concentration ellipses (CE) corresponding to each mass bin. 

At high-redshift $z=15$ (not shown), LVs accumulate at the $a \sim b \sim c$ corner area, as expected for a spherical configuration. 
Then, the three shaded triangles  are progressively filled up. 
We observe that the low-mass LVs tend to be the first to evolve, and the massive ones show a delay in their evolution.  This is the shapes of the LVs where central galaxies with low $M_{\rm vir}^{z=0}$  are going to form evolve first in the axis-ratio plane.

 Concentration ellipses carry out information on the particular evolution trends of the mass groups and   component types (DM or CB). 
Globally, DM-LVs evolve towards low $(c/a)_{\rm DM}$ and high $p_{\rm DM}$ configurations. Their 
evolution is rather similar among the different mass groups, and it results into an overall correlation among the $(c/a)_{\rm DM}$ and $(b/a)_{\rm DM}$ ratios, as the CEs show.

Baryon physics adds dispersion to the CB-LV evolution (right column). In this case, massive CB-LV evolution in the $z=1.0-0.1$ interval is detached from that of the other mass groups. At $z=1$, their CE lies in the  triaxial-shape triangle (shaded in grey) at high  $(c/a)_{\rm CB}$ 
values. As in the DM-LV case, evolution carries them, as well as the LVs in other mass groups, towards lower $(c/a)_{\rm CB}$ and higher $p_{\rm CB}$ values, even  at $z=0.5$ this segregation persists, with   massive LVs  still showing a low prolateness ($p_{\rm CB}< 0.5$). 

In addition, a comparison between the $(c/a)_{\rm DM}$  and the $(c/a)_{\rm CB}$ values reveals that, for low $c/a$ values, the $(c/a)_{\rm CB}$ ratio tends to be lower than   $(c/a)_{\rm DM}$, with the lowest $c/a$ ratios being reached by these CB-LVs at $z=0$. Remarkably, this is also true at high $z$, with flattened CB structures appearing early (see e.g. Fig.~\ref{fig:LVevol-plane}, $z=1$ panel, right versus left columns).

In fact, the relation between the shape parameters of the DM-LVs and CB-LVs varies with redshift (see Fig.~\ref{fig:epT-DM-CB}).
While at $z= 0$ few  structures reach $e_{\rm DM} > 0.8$ or 
$p_{\rm DM} > 0.8$, their CB-LV counterparts do to a larger extent. Therefore, in the case of far-from-sphericity structures, the CB-LV variant reaches more anisotropic configurations than their DM-LV counterparts, with the difference increasing as evolution proceeds. 
These differences 
come very likely from the dissipative character of the gas collapse to a structure with low $(c/a)_{\rm CB}$ axis-ratio, regardless of the value of its triaxiality $T$.

We use cumulative distribution functions (CDFs) to quantify the LV shape  evolution. Panels (a) and (b) in Fig.~\ref{fig:LVevol-CDF-ep}  show CDFs of the  $e_{\rm DM}$ and the $e_{\rm CB}$ ellipticity parameters, respectively. Similarly, panels (c) and (d) refer to the  $p_{\rm DM}$ and the $p_{\rm CB}$ prolateness parameters, respectively. 
The CDF evolution of both shape parameters shows generic  patterns that are the same irrespective of   the mass component used to trace them:  median  values increase (i.e., LV  ellipticity  and prolateness increase).  The $p$ distributions widen, while  $e$ CDFs dispersion remains roughly constant.

It turns out that prolateness widening, or ellipticity changes, occur to different extents depending on the LV  mass (or $M_{\rm vir}^{z=0}$) group.
Second row panels in Fig.~\ref{fig:LVevol-CDF-ep}
illustrate the effect of the LV shape at $z=0$  on the central galaxy halo mass  $M_{\rm vir}^{z=0}$ (recall the tight, extremely low dispersion correlation between halo mass and $M_{\rm LV}$). 
We confirm  that low-mass DM-LVs (orange lines) evolve preferentially into extremely prolate shapes at $z=0$. 
Regarding CB-LVs, while a few ($\sim 22\%$) massive LVs (magenta line)   become  extremely prolate ($p_{\rm CB}>0.5$), most massive CB-LVs (a $\sim 59\%$) are still far from being prolate  (i.e., they have $p < 0.2$) at $z=0$. 

Similar qualitative results are obtained when  CB-LV shapes are analysed using the $e$ and $T$  parameters.  We find that a fraction of low-mass haloes as  high as more than  $80\%$ are formed within CB-LVs with $T_{\rm CB} > 0.7$ at $z=0$. Conversely, the fraction of massive  haloes at $z=0$ formed in LVs with $T_{\rm CB} > 0.7$ is lower, and even lower than that of massive haloes  in the $0.3<T_{\rm CB}<0.7 $ group.

The relationship between  the LV mass and the DM-LV shape is also very strong, when quantified using the two-sample Kolmogorov-Smirnov (KS) test and comparing the massive and low-mass groups, see p-values  on each plot of Fig.~\ref{fig:LVevol-CDF-ep}.  
Note the robustness of the results when shape evolution is studied either with the $p$ or $e$  parameters, reflected in the very low p-values.

These findings pose 
 the question of why the shape of a LV at $z=0$  affects the $M_{\rm vir}^{z=0}$ of the central galaxy they host.   The answer can be found  in the last paragraphs of Section~\ref{sec:IndLVShape}.

\subsection{Freezing out of eigen-directions and shapes}
\label{sec:freezing-times}

As previously mentioned, at a given time the size of the principal axes of inertia do not vary much with time, i.e. they become frozen, see Fig.~\ref{fig:prinaxesevol}. In order to quantify these freeze out timescales, as in \citet{Robles:2015}, we define a fractional threshold $f_a$, i.e. a fraction of $a(t)$ such that  if $t_{\rm U} \ge t_{f, a}$,  $\Delta a(t) \equiv \frac{\mid a(t) - a(t_{\rm U})\mid}{a(t_{\rm U})}  \le  f_a$, where $t_{\rm U}$ is the age of the Universe and $t_{f,a}$ is the freeze out time of the major axis $a$. Similarly, we define $t_{f, b}$ and  $t_{f, c}$. We take $f_{a,b,c}=0.1$. Then, we calculate $\tfmin$  and $\tfmax$, the minimum and maximum values of the three freeze out times. 

In a similar fashion, a threshold angle $\delta A_i$ is defined such that if $t_{\rm U} \ge t_{\delta A_i}$, $A_i(t) \le \delta A_i$, 
where $t_{\delta A_i}$ is the time when the eigen-direction $i$ becomes frozen, as illustrated in Fig.~\ref{fig:angAievol}. Then minimum $\tdAmin$ and maximum $\tdAmax$ freeze out timescales are calculated. Note here that once one of the eigen-directions is fixed, the remaining two become frozen at approximately the same time. We set  $\cos (\delta A_i) = 0.9$. 

 The previous definitions refer to the case when all the particle types included in the LV are considered. When we refer to DM-LVs or CB-LVs, similar  definitions apply, but relative to the corresponding r-TOI. See Table~\ref{tab:Glossary} for the freezing out timescale notation.

 Next, we highlight the interest of  studying  environment freezing-out timescales in the context of  galaxy formation.
 The freezing out timescale  $\tdAmaxDM$  of a given DM-LV marks the  moment in the cosmic evolution when  its principal directions become frozen (within $10\%$).
 From this moment on, 
we can consider that the DM-LV spine has been set, 
i.e., the migrant mass flows of a given DM-LV have  fixed directions (within $10\%$).  
This may have consequences on galaxy properties, see Section~\ref{sec:GalProp-Lvtmaxmin}.

It is worth noting that, for the CB-LV, reaching their $\tdAmaxCB$ implies the end of other baryonic  processes beyond  their DM-spine emergence, see Section~\ref{sec:IndLVShape}. 
 
On its turn,  $t^{\rm max}_{f, \rm DM}$  marks the moment when the three DM-LV principal axes, $a_{\rm DM}$, $b_{\rm DM}$, and $c_{\rm DM}$, stop changing  (also within $10\%$).
Any anisotropic mass rearrangement induces a change in their values. Thus, their freezing-out can be regarded as the moment when important anisotropic mass density reconfigurations of the DM-LV constituent particles have ceased.  Should this happen at a later time, the forming galaxy would have had a longer period of time to be fed from  its surrounding mass distribution. 

\begin{figure}
\centering
\includegraphics[width=\columnwidth]{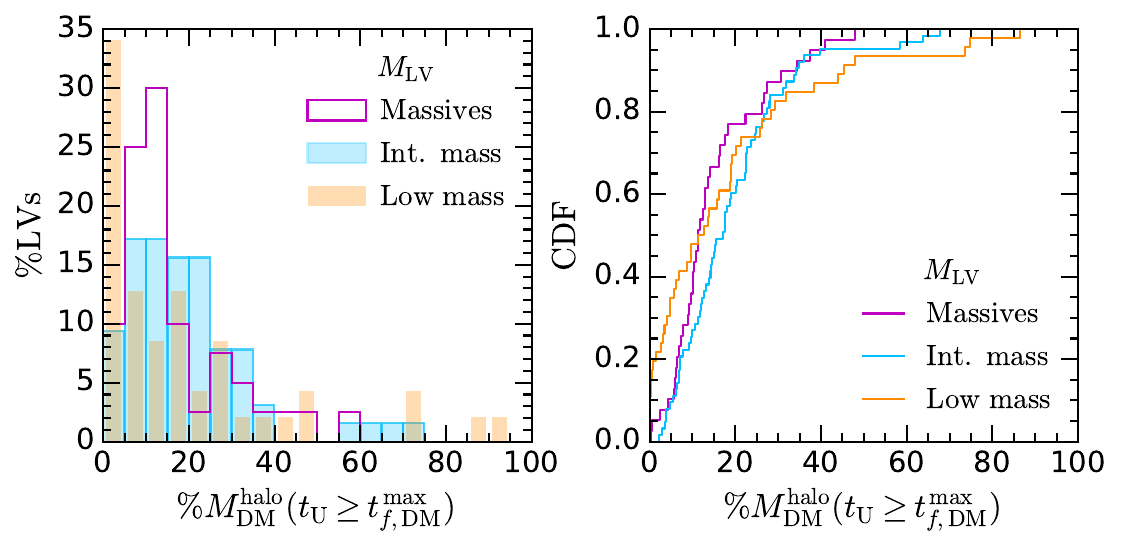}
\caption{Left: Histograms for the fraction of the halo DM mass accreted after its LV reaches $\tfmax$ binned by the halo mass. Right: The corresponding CDFs. }
\label{fig:hist-DM-accre-tfmax}
\end{figure}

To confirm that mass feeding decreases severely after $\tfmaxDM$, in Fig.~\ref{fig:hist-DM-accre-tfmax}  we show  histograms of the fraction of the galaxy DM mass that has been accreted after their  shape parameters $a_{\rm DM}$, $b_{\rm DM}$ and $c_{\rm DM}$ of its corresponding DM-LV stop changing at $t_{\rm U} = \tfmaxDM$.  We see that the fraction peaks at low \% values, particularly so for the low mass  ($\sim 5\%$) and large mass ($\sim 15\%$) bins.  The right panel depicts the corresponding CDFs, where these results are more clearly reported. 

We caution the reader that $\tfmaxCB$ values, calculated from CB particles, imply the end of CB mass flows, and, moreover, the end of baryonic particle type changes through the already mentioned heating/cooling.

In the process of CW setting at high redshift, a common situation is that a DM caustic appears and gains mass because of mass flows towards it (i.e., the so-called migrant mass flows, as previously  mentioned).
For instance, walls grow from mass flows from the nearby mass distribution.  As a result, infra-dense volumes expand and become voids.
Also,  filaments form at the junction of walls (or within them) and grow at the expense of mass in walls.
 In a similar manner clumps grow too. 
A common scenario is that the DM-LV spine is set first, and mass keeps on flowing towards DM-spine elements, and so $\tfmaxDM$ comes after $\tdAmaxDM$. 
 

 \subsubsection{Statistical results of the freezing-out timescales}
 \label{sec:tmaxmin-H}

\begin{figure*}
\centering
\includegraphics[width=\columnwidth]{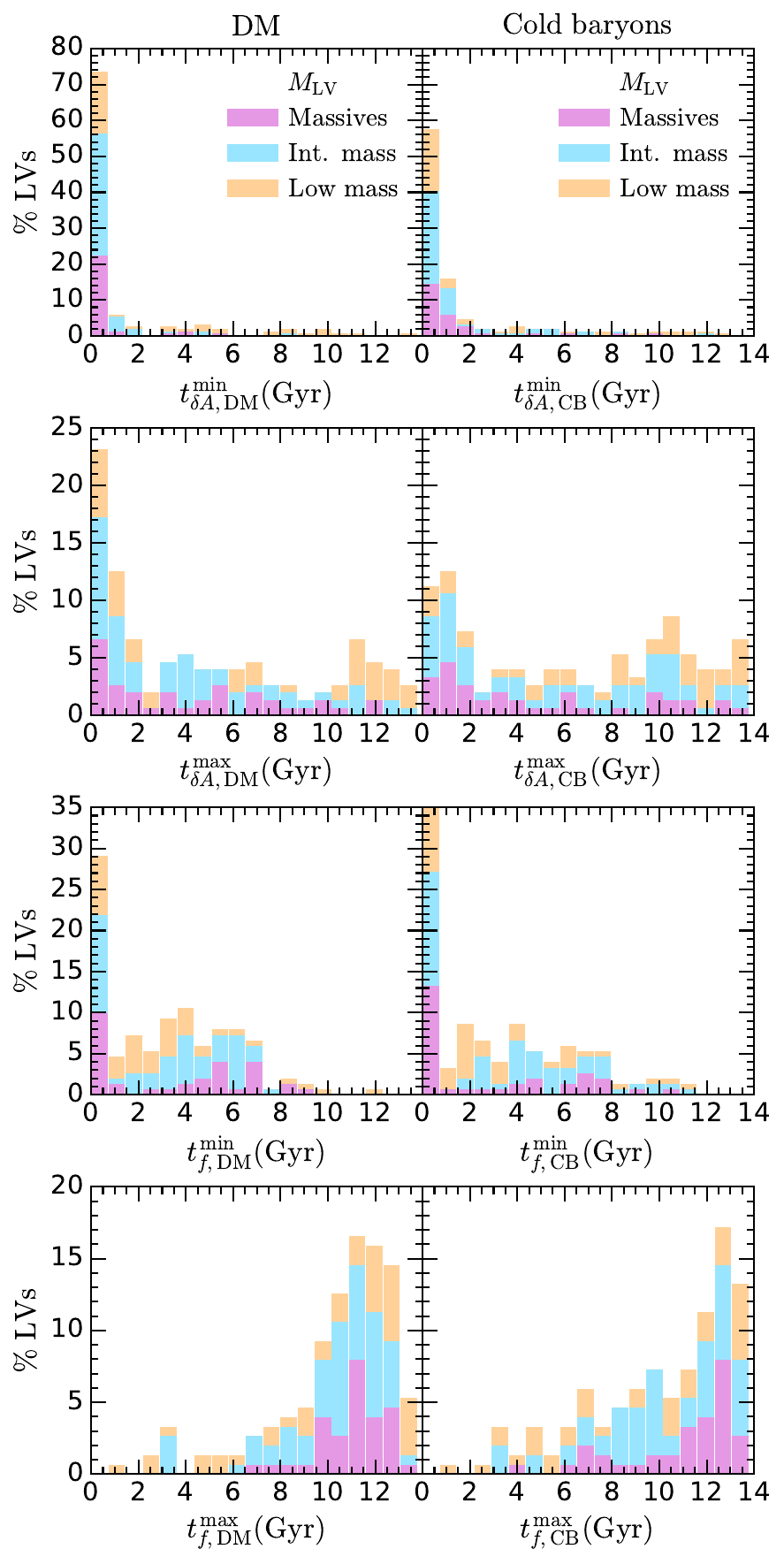}
\includegraphics[width=\columnwidth]{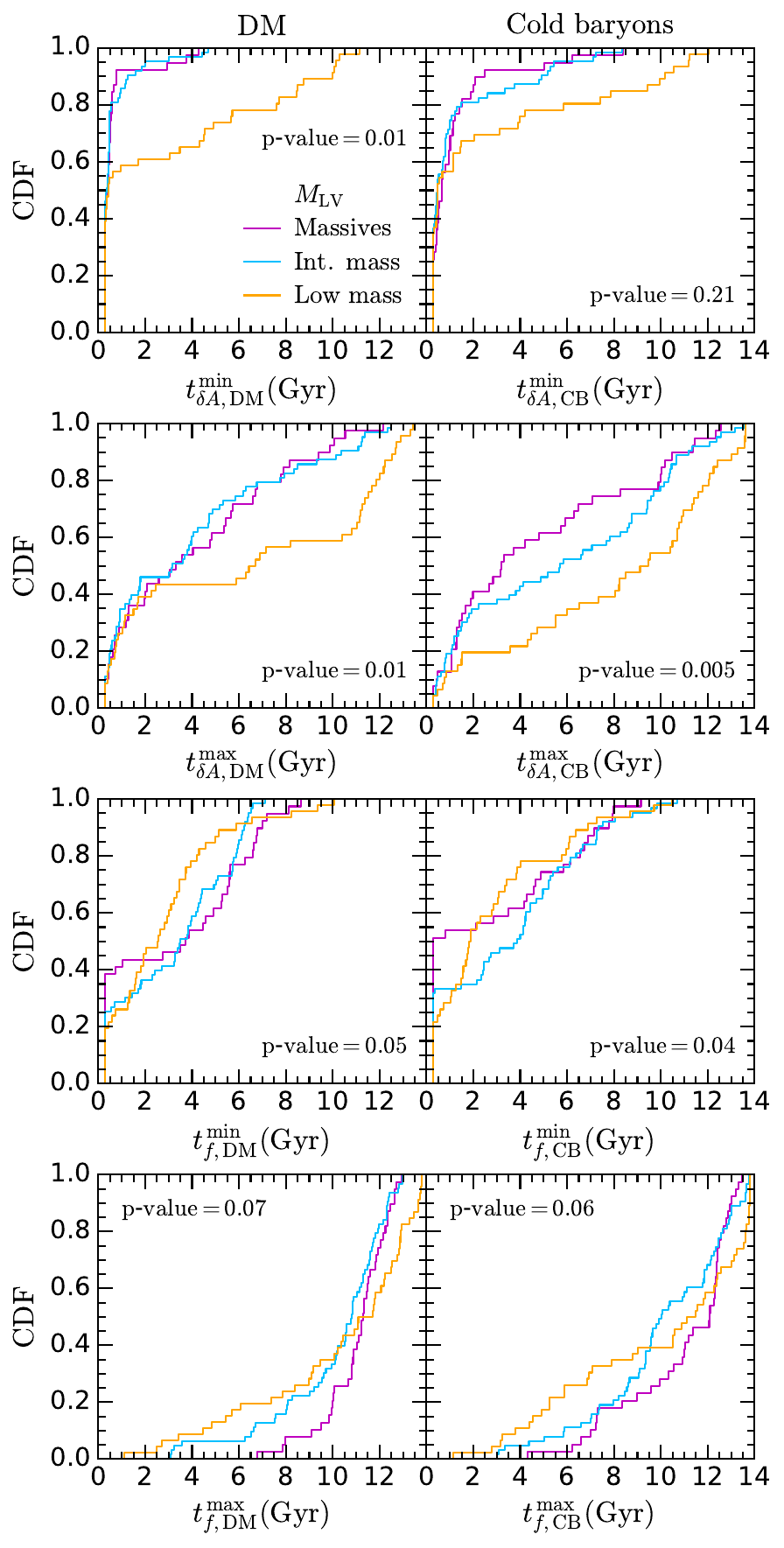}

\caption{Histograms and CDFs for  $\tdAmin$, $\tdAmax$, $\tfmax$, and $\tfmin$
defined with $\cos (\delta A_i) = 0.9$ and $f=0.1$ respectively.
for the DM (first and third columns) and cold baryon components (second and fourth columns). The  p-values of the two-sample Kolmogorov-Smirnov (KS) test correspond to the difference between the massive and low-mass LVs. 
}
\label{fig:Htmax-tmin}
\end{figure*}

 Histograms and CDFs for the freezing out times for both the DM-LV (left panels) and the CB-LVs  (right panels) are given in Fig.~\ref{fig:Htmax-tmin}. The first two rows show the freezing out timescales of the eigen-directions, while  the third and fourth rows correspond to those of the principal axes. 
 We note that one of the eigen-directions of the LVs in our sample is frozen very quickly ($80\%$ of them during the first Gyr of the Universe's  age, see distributions  in the first row of Fig.~\ref{fig:Htmax-tmin}). 
 The distributions of skeleton emergence times is bimodal (second row). About $40\%$ of DM-LVs (and $30\%$ of the CB-LVs) have their skeleton completely formed before $t_{\rm U} \sim 2 \Gyr$.
 In contrast, the  age distribution of the skeleton emergence  in the remaining LVs is rather flat until $z=0$ for DM-LVs, and has a small bump for CB-LVs.   
 
 Let us now analyse the distribution of the principal axes freezing-out timescales.
 A tendency for one of the  principal axes to stop changing very early  is also apparent in Fig.~\ref{fig:Htmax-tmin},  third row. After that, LVs are compressed in one direction and stretched in the other.  They represent the $30\%$ and $35\%$ of the sample for DM- and CB-LVs, respectively, while the rest of the LVs have their $\tfmin$ spread over the first $\sim7$ or
 $8 \Gyr$ of the Universe's age. 
 The distributions of $\tfmax$ differ quite a lot from that of $\tfmin$ as they peak at $t_{\rm U} \sim 10.5  \Gyr$  for DM-LVs and accumulate from $t_{\rm U} \sim 12 \Gyr$  onward in the case of CB-LVs\footnote{Recall that in the case of CB-LVs, not only mass flows but also particle type changes play a role at determining $\tfmaxCB$.}, see bottom panels of Fig.~\ref{fig:Htmax-tmin}. 
 Anisotropic rearrangements of CB-mass as well as particle type changes in CB-LVs occur up to $z=0$.
 Comparing with the plots in the second row, we see that, statistically, anisotropic mass rearrangements continue for a long while  after the DM-LV skeleton is set (fixing migrant mass flows´ directions).

 

 \subsubsection{Effects of the LV mass and particle component  on the freezing-out timescales}
\label{sec:tmaxmin-LV-MassComponent}
 
The effects of the LV mass and particle component (DM and CB) in the freezing out timescales can be inferred from the CDFs in the third and fourth columns of Fig.~\ref{fig:Htmax-tmin}. 

\paragraph*{Freezing-out timescales of the eigen-directions.- }

The skeleton emergence is retarded in low-mass LVs relative to more massive ones. This effect is more pronounced when CB-LVs are analysed than when  DM-LVs are considered, see plots in the second row. This is, the DM skeleton of
massive and intermediate-mass DM-LVs emerges earlier on. 
This effect of the LV mass on  $\tdAmaxDM$ for DM-LVs is a consequence of dynamical evolution being faster in denser environments (see Eqs.~\ref{eq:DenZAppro} and \ref{eq:landscape_height}),  as are those of  massive halo formation sites, remember the very tight correlation between halo mass and $\MLV$. 

As for the particle component effect ($\tdAmaxDM$ CDFs on the third column compared to $\tdAmaxCB$ CDFs on the fourth column), we see that it takes  longer for the CB-LV component to complete its skeleton formation than to DM-LVs, particularly so for $t_U <$ 2 Gyr, see the corresponding histograms on the left, and for the low-mass bin CB-LVs.   
Indeed, the difference among DM- and CB- CDFs is  most important for LVs in the low-mass bin.
Probably due to SNe and BH effects (such as hot gas outflows), and other heating processes,  active at high/intermediate redshift for these LVs hosting low-mass galaxies, see Section~\ref{sec:PriDirecEvol}.
Massive objects are dominated by the DM  as driver of their mass assembly, so that they have these sub-resolution processes quenched from earlier on.
At $t_{\rm U} = 8 \Gyr$, the skeletons of only 40\% (60\%) of the CB- (DM-)LVs 
 in the low-mass bin have been formed, in contrast to 75\% (90\%)  of CB-(DM-)LVs in
  the massive bin.

The effect of the LV mass is also remarkable in $\tdAmin$ (first row panels), where, again, it takes longer for LVs in the low-mass bin to have one eigen-direction frozen, while all the massive LVs have at least one frozen eigen-direction at $t_{\rm U} = 2 \Gyr$  (DM-LVs) or $4 \Gyr$ (CB-LVs).

\paragraph*{Freezing-out timescales of the principal axes.- }

In this case, the freezing out  timescales affect the final halo mass distribution. Indeed,
the three principal axes of the LV massive bin  (or, equivalently, the $\mvirzlow$ mass bin)
are  frozen later on up to
$t_{\rm U} \sim  11 \Gyr$  (or $ \sim  12 \Gyr$  when the  analysis is made using CB-tracers).  After that, the differences are inverted, see fourth row panels. 

The physics behind the behaviour up to $t_{\rm U} \sim 11 \Gyr $  for DM-LVs (or $12 \Gyr$ for CB-LVs) is as follows: i) the larger the time interval along which DM mass flows are active (i.e., the later $\tfmaxDM$),  the more likely the corresponding central galaxy has a large halo mass, and ii) the  mass of the central galaxy and that of the LV are highly correlated, see Fig.~\ref{fig:MstarMLV}, and especially $\mvirzlow$ and $\MLV$.
As for the  change in behaviour after this first period, a possibility is that the DM mass budget available to be assembled into the host galaxy halo (assembly that occurs through migrant flows) is exhausted around  massive haloes, while this is not the case for the lower mass hosts.

Quantitatively, at $t_{\rm U} \sim 10  \Gyr$ only $\sim 30\%$ of DM-LVs in any mass bin have their three principal axes frozen. For the remaining DM-LVs, anisotropic mass rearrangements do occur up to $t_{\rm U} \sim 13 \Gyr $ for massive and intermediate mass DM-LVs (or central galaxies with large or intermediate $\mvirzlow$), while DM mass flows keep occurring for another Gyr  for those in the low-mass bin. In the case of the CB-LVs at $t_{\rm U} \sim 10 \Gyr$, the fractions are slightly higher, and their evolution is  qualitatively similar.

\subsubsection{Eigen-directions and principal axes freezing-out ordering}
\label{src:Ordering-fo-LVs}

\begin{figure}
\includegraphics[width=\columnwidth]{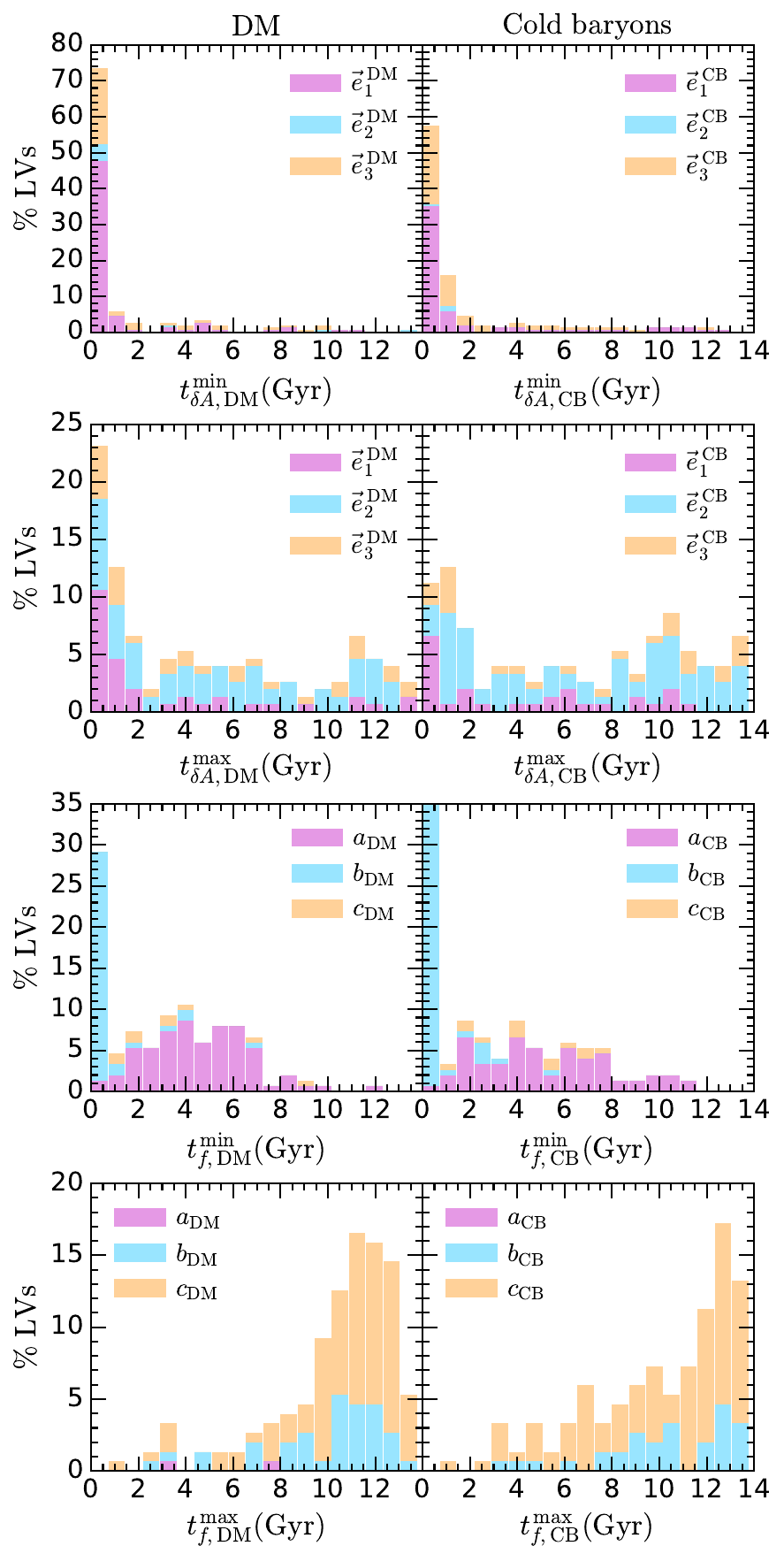}
  \caption{Histograms for the relevant  timescales of the LV sample, split according to which of the eigen-directions (first and second row) or the principal axes (third and fourth row) freezes out first (min) or last (max). 
  Timescales have been calculated using either DM  (left column) or CB particles (right column) sampling of each LV.}
\label{fig:FirstToFrozen-LV}
\end{figure}

An interesting question  is which  among the three principal directions is the first to freeze out. Figure~\ref{fig:FirstToFrozen-LV} addresses this question in the first row panels by showing $\tdAmin$ histograms binned by this eigen-direction.  
We see that irrespective of how the LV is analysed using DM or CB particles, most LVs have their $\vec{e}_1$ eigen-direction fixed first. 

Regarding the last eigen-direction to freeze-out, panels in the second row of this figure, $\tdAmax$ histograms,  inform us, 
that for most of the LVs whose skeleton emerge after the first $2 \Gyr$ of the  age of the Universe, the $\vec{e}_2$ and $\vec{e}_3$ eigen-directions (light blue and orange, respectively)  are the last to be fixed. Specifically for DM-LVs, when their last eigen-direction to freeze out is $\vec{e}_1$, the DM skeleton emergence occurs within the  first $2 \Gyr$ of the  age of the Universe.

In the third row panel, histograms are classified according the principal axes  $a, b,$ and $c$ freezing-out order (magenta, light blue, orange bars, respectively). 
 In most  cases, it is the $b$-axis, for both DM-LVs and CB-LVs, the first to be fixed at very early ages. This was expected, see Section~\ref{sec:IndLVShape}.

Finally, bottom row panels show details of the end of anisotropic mass reconfigurations in LVs. They indicate that either for DM- or CB-LVs, the last principal axis to stop changing is, in most cases, the minor axis $c$  (i.e., that associated to the most compressive  mass flows, with an important component along the $\vec{e}_3$ principal direction), and it is never the major axis $a$. 
We also see that the end of CB-mass rearrangements is delayed relative to that of DM.  
For a possible explanation of such a behaviour see Section~\ref{sec:IndLVShape}. 

It is worth noting  the segregated character of the results obtained for the principal axes (panels in the third and fourth rows). In Fig.~\ref{fig:FirstToFrozen-LV}, we see that for $30\%$ ($35\%$) of  DM- (CB)-LVs, the first principal axis to freeze out is $b$.  For another $\sim$  third of them, $b$ is the last principal axis to stop changing, preferentially at late times ($t_{\rm U} >8 \Gyr$). The other two principal axes are either the  first to freeze-out ($a$), up to $\sim8 \Gyr$, or stop changing the last ($c$) for a high percentage of cases. 

This particular role of the minor axis $c$ in LV deformation is likely related to the specific role that flattened structures play in the  CW evolution.
Indeed, $c$ is linked to the compressional component of the tidal CW cosmic field, responsible for  shaping walls,  where two voids meet after having expanded (i.e., matter compression pointing towards the forming wall placed at the void boundaries).  
The physical relevance of voids as the spatial organisers of the CW structure is becoming more clear since \citet{Icke1984}  seminal work  until the latest developments \citep[see e.g., ][]{Kugel2024}.

The study of the effect of the freezing out timescales axis and direction ordering on LV masses is postponed until Section~\ref{sec:f-oOrderingEffects}, as their physical motivation comes from the  effects on the host galaxy mass, which is tightly correlated to the LV mass.


\subsubsection{Relations among LV freezing-out timescales and shape parameters}
\label{sec:LV-Tep-tmaxmin}

\begin{figure}
\centering
\includegraphics[width=\columnwidth]{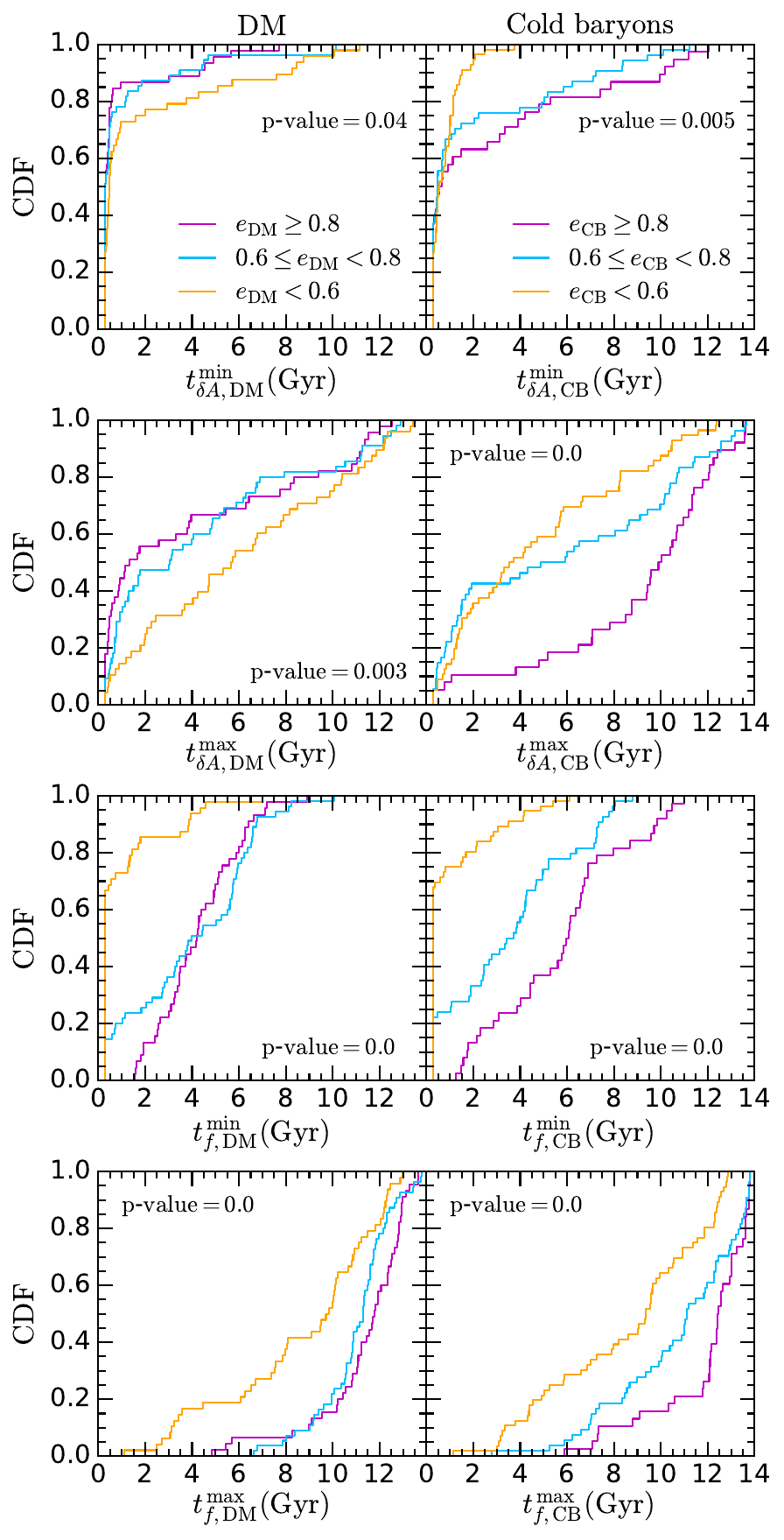}
\caption{CDFs for the freezing out timescales, binned by the LV ellipticity parameter at $z=0$ of the 
 DM-LVs (left panels) and  cold baryon LV component (right panels). First and second rows: eigen-directions  timescales. Third and fourth rows: principal axes timescales. The  p-values of the two-sample KS test correspond to the difference between the first and third ellipticity bins.  }
\label{fig:CDF-LV-pr}
\end{figure}

To examine if there is any  relation among LV shapes and  freezing-out  timescales, we binned the latter   
by the $e$ parameter at $z=0$ in Fig.~\ref{fig:CDF-LV-pr}. 
We first note that,  at $z=0$,  the sample of low $e_{\rm DM}$ and $e_{\rm CB}$ in the DM-LVs and CB-LVs, respectively,  does not consist of the same individual LVs (see Fig.~\ref{fig:epT-DM-CB}). 
Indeed, we see in this figure that the correlations among the ellipticity  $e_{\rm CB}$ and $e_{\rm DM}$ parameters is very low. The same is true for the $p$ parameter and for the triaxiality $T$ (not shown).

The effect of the LV shape as measured by the ellipticity parameter $e$  on the principal directions freezing-out timescales is shown in the first and second  rows of Fig.~\ref{fig:CDF-LV-pr}. 
In the second row of DM-LVs we see that  the skeleton of $\sim$ half of the  highly elliptical ones ($e_{\rm DM} \ge 0.8$, magenta lines)  emerges very early and before that of those with lower values of the $e_{\rm DM}$ parameter.
The effect is reversed, however, when CB-LVs are considered (right panels in the same row). In this case, it takes longer for the spine of highly elliptical CB-LVs (high $e_{\rm CB}$, magenta line) to emerge than for 
that of those with low $e_{\rm CB}$. We also see that this reversed behaviour (of DM-LVs as opposed to the CB-LVs) mostly affects the high $e$ LV bin when compared to the low $e$ LV bin of the DM- and CB-LVs (different samples), while those in the intermediate $e$ value bins (light blue) have  less dissimilar $\tdAmax$ CDFs. 
Presumably, this important delay in  the CB-spine emergence relative to that of DM, in extremely elliptical DM- versus CB-LVs, is related to the already mentioned internal energy and pressure sources effects, absent in DM LVs.  This affects LV shapes so that the DM and the CB shapes are not correlated at $z=0$,  see Fig.~\ref{fig:epT-DM-CB}.

Regarding the principal axes freezing out timescales, the distribution of $\tfmin$ is  bimodal. 
We see that  one of the principal axes of  $\sim 65\%$ of the  low ellipticity ($e_{\rm DM}< 0.6$) DM-LVs (orange line on the third row  panels in Fig.~\ref{fig:CDF-LV-pr}), freezes as early as $t_{\rm U} < 1 \Gyr$, and that of $\sim 95\%$ of them at $t_{\rm U} < 4 \Gyr$, and similarly for CB-LVs.  
Recall that once this happens,  LVs are compressed in one direction and stretched the other. 
Conversely, one of their principal axes of 
all high $e$ DM-LVs (magenta line) freezes at  $t_{\rm U} \sim 8 \Gyr$  (see third row panels), and later on for the CB-LVs. 

Additionally, CDFs of $\tfmax$ (bottom row panels) clearly indicate that it takes longer for high $e$ LVs ($e \ge 0.8$), to have their three principal axes frozen, compared to those with low $e$ values, for both  DM- and CB-LVs, with the latter showing a stronger effect.

Similar results are obtained when examining CDFs for the  $p$ parameter. 
We note that LVs for which  one of their principal axes is fixed very early  (the $b$-axis, necessarily freezes at its  initial value, i.e.,  $b \sim \sqrt{1/3}$) 
 cannot acquire a prolate shape by deformation. Indeed, prolateness would demand a shrinkage of the $b$-axis. These LVs have $T < 0.7$, and they include the most massive LVs in the sample.

\section{The role of LV evolution in  galaxy properties}
\label{sec:RolLVHost}

In Section \ref{sec:LV-shape-evol}, we briefly commented on the implications  that the shapes LVs acquire around forming galaxies could have on galaxy properties.  We pointed out that LV shapes are interesting in this respect, both as trackers of large-scale forces acting on the LV mass, and  because of the geometric limitations a LV shape imposes on the degrees of freedom of material accreting  onto a  (proto-)galaxy. In fact, we have shown and justified the effect of the DM-LV shape  on the corresponding central galaxy  halo mass at $z=0$,  $\mvirzlow$.

 Then, at the beginning of Section~\ref{sec:freezing-times}, we underlined the interest of  statistical results of  LV shape and orientation  freezing out timescales to study  environmental effects on galaxy formation. 
We discussed  the mutual effects between freezing out timescales and the $\mvirzlow$ of the central galaxy.
Also, in Section~\ref{src:Ordering-fo-LVs}, the possible effects of eigen-directions and principal axes freezing-out ordering  were examined.

In view of the foregoing, in this section  we  analyse the  role  of these  LV characteristics in the determination galaxy properties, namely  shape evolution,  freezing out timescales, and their respective ordering. 
Aside from the central galaxy halo mass, $\mvirzlow$, previously analysed, the following galaxy properties are explored in this section: 
(i) the total  stellar mass of the galaxy, $\mstar$, and 
(ii) the  $\kappa^{\rm rot}$ morphological parameter which is the fraction of kinetic energy involved in ordered rotation \citep{Sales:2012}.  


\subsection{Correlations among galaxy properties and LV shape}
\label{sec:GalProp-Lvshape}

In this section, we study whether or not LV shape  affects the stellar mass and kinematic morphological parameter of the galaxies they host, by means of the LV  axis-ratios, as well as the $T$, $e$, and $p$  shape parameters. 

To begin with, there is a clear correlation between the galaxy stellar mass and the total mass of the LV that fed it, see
Fig.~\ref{fig:MstarMLV} (or equivalently, as previously mentioned, a correlation with the central galaxy $\mvirzlow$). 
Thus, this figure is equivalent to  the stellar/halo mass correlation in galaxies \citep{Wechsler:2018}. Given that the host galaxy halo  masses $\mvirzlow$ are related to  the  shapes of the LV that fed them (see Section~\ref{sec:LV-shape-evol}), some effects can be expected. 
To quantify them, CDFs binned in the above-mentioned galaxy properties  have been calculated. We discuss these results below.


\subsubsection{Galaxy stellar mass}
\label{sec:ShEffMstar}

Here, we aim to answer the question if the LV shape  affects the stellar mass of the galaxy that the LV fed on.

In Section~\ref{sec:ShapeEvolStat}, we have   shown that those DM- or CB-LVs evolving into highly prolate or elliptical shapes host preferentially low $\mvirzlow$ central  haloes, see Fig.~\ref{fig:LVevol-CDF-ep}. A result qualitatively  similar has been found for the $T$ triaxiality parameter (not shown).
The mentioned correlation between $\mstar$ and $\mvirzlow$ implies an effect of the LV shape on the former. To make sure that this is the case, the CDFs of the shape parameters $p$, $e$,  and $T$ binned in $\mstar$ have been calculated. The $\mstar$ bin limits have been determined by those used for $M_{\rm LV}$ in  Fig.~\ref{fig:MstarMLV}, namely $1.5\times 10^9 M_\odot$ and $3\times 10^{10} M_\odot$. 
In this way, the identities of the LVs in each of the 3 stellar mass bins remain approximately the same as those in the $M_{\rm LV}$ mass groups. 
 
 The resulting CDFs for the  $e$ and $p$ shape parameters are quite similar to those depicted in the second row plots of Fig.~\ref{fig:LVevol-CDF-ep}, CDFs describing the LV shape effect  on the host halo mass at $z=0$. Thus, the same related comments in Section~\ref{sec:ShapeEvolStat} apply here, but relative to the LV shape (parametrized using $p$ and $e$) effect on $\mstar$. 
These can be summarized as follows. 
 Central galaxies with low  $\mstar$ form preferentially within prolate-like or elliptical-like  shaped LVs.

Concerning triaxiality, a segregation appears between galaxies with large and low $\mstar$  in the sense that the latter more likely  forms  within DM-LVs with a higher $T_{\rm DM}$ value (i.e., more triaxial) than the former. E.g., only  $15\%$ of low-mass galaxies form in DM-LVs with $T_{\rm DM} < 0.6$. This segregation also appears  in the CB-LV case, where  $60\%$ of massive galaxies live within  CB-LVs with $T_{\rm CB} < 0.7$, in contrast to only $35\%$ of galaxies in the low-mass bin.




\subsubsection{Galaxy $\kappa^{\rm rot}$ kinematic morphological parameter}
\label{sec:ShapeOnKrot}

The  $\kappa^{\rm rot}$ parameter  was originally defined as the fraction of the stellar kinetic energy of a given galaxy  coming from  ordered rotation \citep{Sales:2012,Du:2019}, namely
\begin{equation}
\kappa^{\rm rot}=\frac{1}{M_\star}\sum_i m_i \left(\frac{v_{i,\phi}}{v_i}\right)^2,
\label{eq:KrotDef}
\end{equation}
 where $M_\star$ is the total stellar mass of the galaxy,  $v_{i,\phi}$ and $v_i$ are the tangential velocity of a stellar particle of mass $m_i$ in the plane of the galactic disc (plane normal to the disc rotation axis), and the modulus of the velocity of this $i$-th  stellar  particle, respectively.
\citet{Sales:2012} showed that the morphology of a galaxy is largely determined by the convenient  alignment of the
angular momentum of baryons being accreted over time, that feed on the proto-galaxy,  with the later instantaneous spin at accretion time. 
Thus, $\kappa^{\rm rot}$ at a given age of the Universe, $t_{\rm U}$, is a measure of how well this alignment has worked out  up to $t_{\rm U}$.
Furthermore, $\kappa^{\rm rot}$ is a kinematic indicator of the morphology of a galaxy, such that a galaxy with $\kappa^{\rm rot} <  0.5$ is considered to be a kinematic spheroid (dispersion-dominated), while galaxies with  higher values are rotation-dominated, 
i.e., discy or can be regarded as kinematic discs   \citep[see also ][]{Correa:2017,Du:2019,Celiz:2025}.

 The definition of $\kappa^{\rm rot}$ can be extended to other galaxy components, such as cold gas and cold baryons, just by making the substitution of stellar particles by CG or CB particles. 
We have calculated the $\kappa^{\rm rot}$  for the galaxies in our sample,
considering either stellar, cold gas or CB  particles\footnote{$\kappa^{\rm rot}$ parameters  have been calculated out of particles within $1.5 \, R_{\rm opt}$, where $R_{\rm opt}$ is the optical radius of the central galaxy at the corresponding redshift $z$.
We have tested that using $2$ or $2.5 \, R_{\rm opt}$ instead  returns the same results. It is also worth noting that the number of galaxies for which $\kappa_{\rm rot}$  can be properly calculated is lower  than the number of LVs (121 versus 151, respectively). This is especially relevant for those LVs hosting low-mass galaxies, as the low number of particles in these galaxies prevents their optical radius calculation.} (hereafter  $\kappa^{\rm rot}_{\star}$,
$\kappa^{\rm rot}_{\rm CG}$ and $\kappa^{\rm rot}_{\rm CB}$, respectively). 
Some relationships among each of them  can be found in Fig.~\ref{fig:Krot-vs-mass}, as well as its variation with respect to the stellar mass within the galaxy optical radius (see Appendix~\ref{app:appendixA} for further details).

 It is worth noting that rotation supported  central galaxies are not very abundant in the EAGLE50 simulation, see Table~\ref{tab:rotators} and CDFs in Fig. \ref{fig:Krot-vs-mass}. 

\begin{figure}
\centering
\includegraphics[width=\columnwidth]{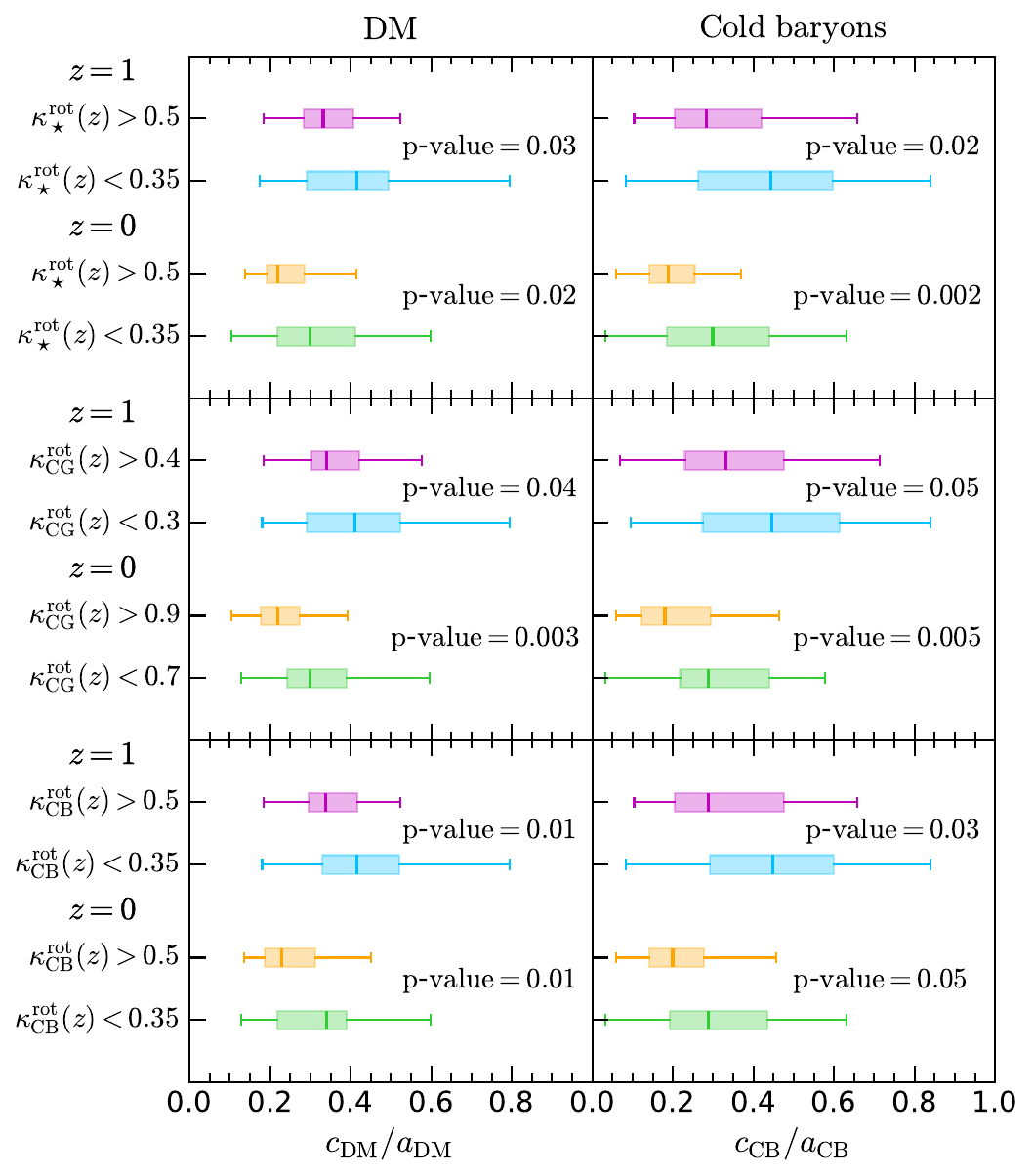}
\caption{Boxplot of the $c/a$ axis-ratio for both DM- (left column) and CB-LVs (right column) binned by $\kappa^{\rm rot}_{\star}$ (first row), $\kappa^{\rm rot}_{\rm CG}$ (second row) and $\kappa^{\rm rot}_{\rm CB}$ (third row) at $z=1$ and $z=0$. Each box is delimited by the first and third quartiles, and the vertical line inside it denotes the median value. The  p-values of the two-sample KS test correspond to the difference between bins at the same $z$.  }
\label{fig:cdf-ca-krot-binned}
\end{figure}

To  look for a potential effect of the LV  shape  on $\kappa^{\rm rot}_{\star}$, $\kappa^{\rm rot}_{\rm CG}$ and $\kappa^{\rm rot}_{\rm CB}$, 
the LV sample has been split  into three groups, see Fig.~\ref{fig:cdf-ca-krot-binned}, e.g. 
for $\kappa^{\rm rot}_{\star}$ the threshold values are 0.5 and 0.35 (corresponding to bins with 35, 49, and 37  galaxies, from higher to lower values,  respectively).
The corresponding boxplots (schematic CDFs) in terms of the $c/a$ axis-ratio for both DM-LVs and CB-LVs are depicted in Fig.~\ref{fig:cdf-ca-krot-binned}. We show results for $z=1$ and $z=0$.

 A first interesting result at $z=0$ is that the $c/a$ axis-ratio medians for rotation-dominated galaxies (in orange) reach  lower values, for any of the three $\kappa^{\rm rot}$ variants,  than for dispersion-dominated galaxies (in green)\footnote{A low $(c/a)_{\rm CB}$ implies  a high  CB-LV ellipticity, regardless of the value of the triaxiality $T$, see Eqs.~\ref{eq:Tdef}, \ref{eq:ShapePDef} and Fig.~\ref{fig:LVevol-plane}.}. 
The effect is slightly more pronounced in CB-LVs concerning median  values, but dispersions are narrower for DM-LVs. 
This can be understood as the result of an increasing thinness of the LV configuration improving the probability of a coherent 
alignment of the angular momentum of baryons accreted over time, that feed on the proto-galaxy. Indeed, the  thinner  the CB-made inertia ellipsoid, the more the motion of 
 its constituent CB particles is confined to two  dimensions (either as a consequence of in-plane motion or of spiral motion in a filament-like structure, i.e., when $b \sim c$).   
 Hence,  their orbital poles become clustered.

To gain insight into the evolution of $\kappa^{\rm rot}$, we have repeated the same analysis over  $c/a$  at $z=1$ (in magenta and light blue).  
We observe the lowering
of the $c/a$ axis-ratio from $z=1$ to $z=0$, in particular in the median values, with the difference being  more remarkable for the CB component, 
probably  a consequence of cooling via gas dissipation only noticeable in this component. 

Overall, our findings show that  for different bins, the higher the $\kappa^{\rm rot}$ parameter for a galaxy (either measured using stars, cold gas or cold baryons), 
the higher its probability of being formed in a CB-LV with a low $c/a$ axis-ratio  
(i.e., highly elliptical CB-LVs).

\begin{figure}
\centering
\includegraphics[width=\columnwidth]{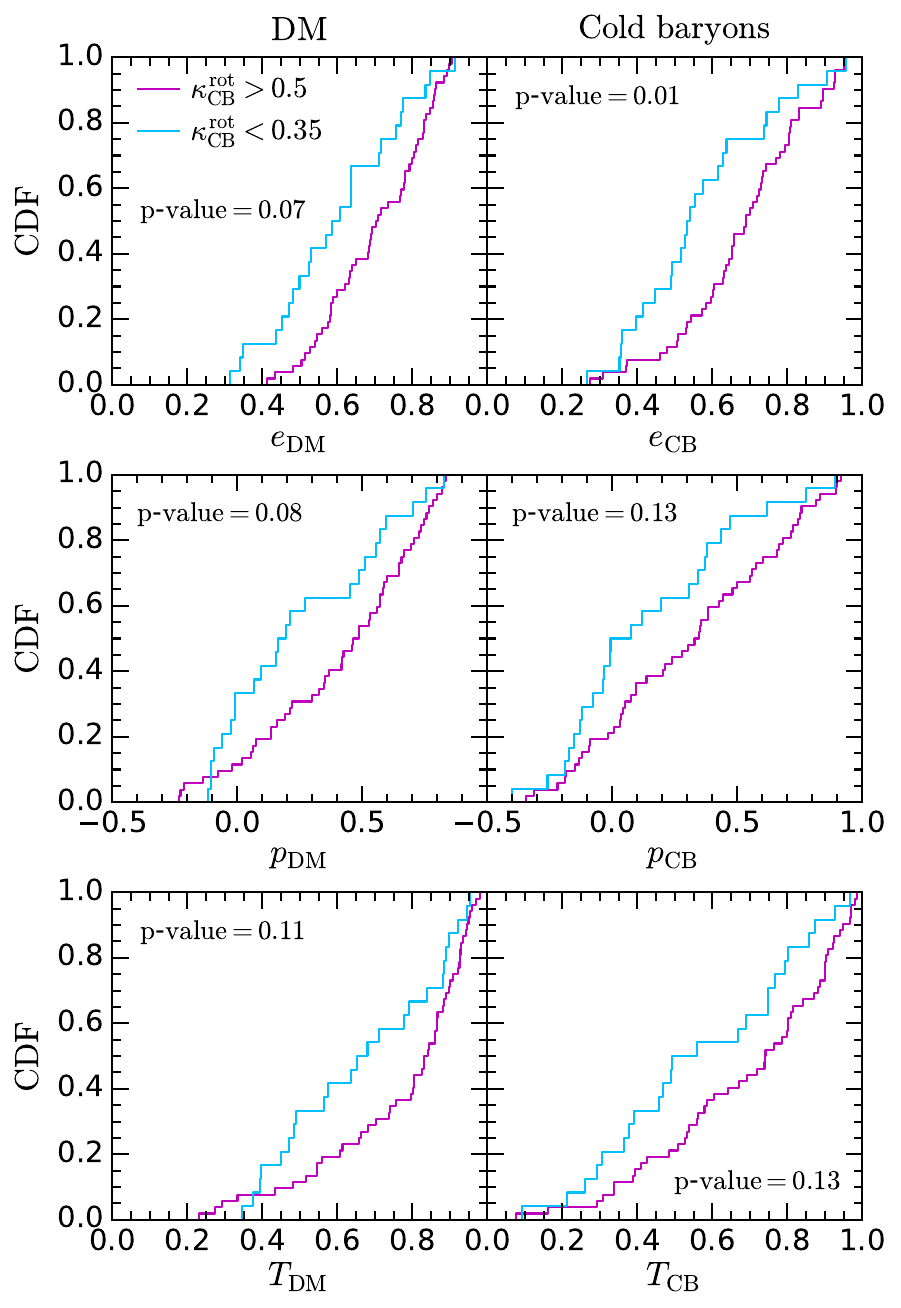}
\caption{CDFs for the   
ellipticity parameter $e$ (top row),  
prolateness parameter $p$ (middle row) and  triaxiality $T$ (bottom row), binned by $\kappa^{\rm rot}_{\rm CB}$.
Left column displays the CDFs for the DM-LVs; right columns refer to CB-LVs. 
The  p-values of the two-sample KS test are also shown. }
\label{fig:epT-CDF-KrotBin}
\end{figure}

The effect of the LV  shape on the $\kappa^{\rm rot}$ parameters can be further quantified  using 
explicit  CDFs, see e.g. Fig.~\ref{fig:epT-CDF-KrotBin}, where CDFs binned by $\kappa^{\rm rot}_{\rm CB}$ are shown. Regarding ellipticity (first row), a tendency appears  when analysing the LVs using both DM and CB components: LVs with $e < 0.6$   harbour a few rotation-dominated galaxies  (magenta line), while they host $\sim 60\% $ of dispersion-dominated systems ($\kappa_{\rm CB}^{\rm rot} < 0.35$, light blue line). 
 Furthermore, while only $\sim5\%$ of DM-LVs and  $\sim 10\%$ of CB-LVs with $e < 0.5$ host  CB kinematic discs 
(magenta line), they have fed on $\sim 25\%$ of kinematic spheroids.

Using $p$ and $T$ parameters provides complementary insight, see Fig.~\ref{fig:epT-CDF-KrotBin}, second and third rows respectively, and Table~\ref{tab:rotators}. 
In this table, we can see that  while the  prolateness of CB-LVs does not seem to affect the fraction of rotators  (roughly one third in each case,  regardless of the  
 $\kappa^{\rm rot}$ variant used to measure them), the prolateness of DM-LVs does affect this fraction. 
 In other words,  rotation-dominated galaxies do not preferentially  accumulate at any $p_{\rm CB}$ interval,  while  
$\sim 50\%$ of galaxies with high $\kappa^{\rm rot}_\star$ or $\kappa^{\rm rot}_{\rm CB}$ (and only $\sim30\%$ of dispersion-dominated) have been fed on by extremely prolate DM-LVs ($p_{\rm DM} > 0.5$). 
In addition, a fraction as high as $\sim 64\%$ of central galaxies with their cold gas in  ordered rotation appear in extremely prolate DM-LVs.
 
It is worth noting that an extremely prolate LV also includes filamentary-like structures (i.e. those with $c,b\ll a$), in which case  particle pole alignments might result from spiraling motion towards the galaxy formation site rather than in-plane motion. 

As for low prolateness configurations,
from Fig.~\ref{fig:epT-CDF-KrotBin} we see that  DM-LVs  with $|p_{\rm DM}| < 0.2$  host  $\sim 50 \%$ of the dispersion-dominated hosts (light blue line) and only $20\%$  of rotationally supported galaxies.  
Regarding the triaxiallity of DM-LVs, most rotators ($70\% - 77\%$) have $T_{\rm DM} > 0.7$, i.e., they are neither triaxial or oblate,  and very few are oblate ($\sim 6\% - 9\%$).

The important point to highlight here is that the shape a LV acquires  along Gyrs of cosmic evolution at Mpc scale,  
seems to affect the fraction of  kinetic energy in ordered rotation that  the central galaxy forming inside the LV shows at $z=0$.



\begin{table}
    \centering
    \caption{Number of rotators, according to the three $\kappa^{\rm rot}$ variants, found in the sample of 121 LVs whose host galaxy optical radius can be calculated  (first row). Fractions of these rotationally supported galaxies are given in different prolateness ($p$) and triaxiality ($T$) intervals. 
    Columns distinguish these fractions according to the $\kappa^{\rm rot}$ variant specified  at the header. Results for both DM- and CB-LVs are listed. }
    \begin{tabular}{cccc}
    \hline
      &  $\kappa^{\rm rot}_\star>0.5$ & $\kappa^{\rm rot}_{\rm CG}>0.9$ &  $\kappa^{\rm rot}_{\rm CB}>0.5$ \\
      \hline
     \# Rotators & $35$ & $36$ & $53$  \\
     \hline 
     $p_{\rm DM}>0.5 $    & $48.6\%$ & $63.9\%$ & $47.2\%$  \\
     $0.2<p_{\rm DM}<0.5 $    & $28.6\%$ & $11.1\%$ & $26.4\%$  \\     
     $p_{\rm DM}<0.2 $    & $22.8\%$ & $25.0\%$ & $26.4\%$  \\
      \hline
     $p_{\rm CB}>0.5 $    & $34.3\%$ & $33.3\%$ & $34.0\%$  \\   
     $0.2<p_{\rm CB}<0.5 $    & $22.8\%$ & $36.1\%$ & $28.3\%$  \\   
     $p_{\rm CB}<0.2 $    & $42.9\%$ & $30.6\%$ & $37.7\%$  
     \\   
      \hline    
     $T_{\rm DM}>0.7 $    & $77.1\%$ & $72.2\%$ & $71.7\%$ \\
     $0.3<T_{\rm DM}<0.7 $  & $14.3\%$ & $22.2\%$ & $22.6\%$ \\     
     $T_{\rm DM}<0.3 $    &  $8.6\%$ & $5.6\%$  & $5.7\%$  
 \\
       \hline
     $T_{\rm CB}>0.7 $    &  $51.4\%$  & $63.9\%$ & $56.6\%$  \\
     $0.3<T_{\rm CB}<0.7 $  &  $40.0\%$  & $33.3\%$ & $37.7\%$ \\     
     $T_{\rm CB}<0.3 $    &  $8.6\%$  & $2.8\%$ & $5.7\%$  \\      
      \hline
    \end{tabular}
    \label{tab:rotators}
\end{table}


\subsection{Relations among  LV freezing out timescales and galaxy properties }
\label{sec:GalProp-Lvtmaxmin}

As explained in Section~\ref{sec:tmaxmin-H},     the freezing out timescale $\tdAmax$  of a given LV can be regarded as the age of the  Universe when the local CW skeleton around the forming galaxy is set in, so that 
the directions of migrant mass flows transporting mass from low-density regions towards singularities become fixed  within  $10\%$. 
On its turn,  $\tfmax$ can be considered as the moment when important mass density reconfigurations of LV constituent particles  (recall that the particle type changes in the case of CB-LVs) cease, limiting in this way important mass exchanges (and  energy transfer in the CB-LV case) between the forming galaxy and its environment.
For these reasons, it can be expected that the properties of a galaxy formed within a given LV depend on the LV freezing out timescales.

In this section, we analyse the possible implications that the freezing out timescales of the LV sample can have on the properties of the galaxies they host. To this end, we study the  $\tdAmin$, $\tdAmax$, $\tfmin$, and  $\tfmax$ CDFs, by splitting these timescales into different subsamples. The same galaxy properties as in Section~\ref{sec:GalProp-Lvshape} have been considered for binning purposes.


\subsubsection{Galaxy stellar mass}
\label{ss-galmasseffects}

For the same reasons explained in Section~\ref{sec:ShEffMstar}, it turns out that the CDFs in the third and fourth columns in Fig.~\ref{fig:Htmax-tmin} are very close to the CDFs for $\tdAmax$, $\tdAmin$, $\tfmax$ and $\tfmin$ binned by $\mstar$. Thus, the comments made in that section on LV freezing out timescales relationships with $M_{\rm LV}$ or $\mvirzlow$, apply here for $\mstar$. We summarize them as follows. Central galaxies with large $\mstar$  tend to form within (i)  LVs with late freezing out  of their three principal axes (except  when the freeze out occurs after $t_{\rm U} \sim 11 -12 \Gyr$, a time interval when low $\mstar$ galaxies form), and (ii) those LVs with an  early setting in of their skeleton.


\subsubsection{$\kappa^{\rm rot}$ morphological parameter} 
\label{sec:KrotCentrals}

\begin{figure}
\centering
\includegraphics[width=\columnwidth]{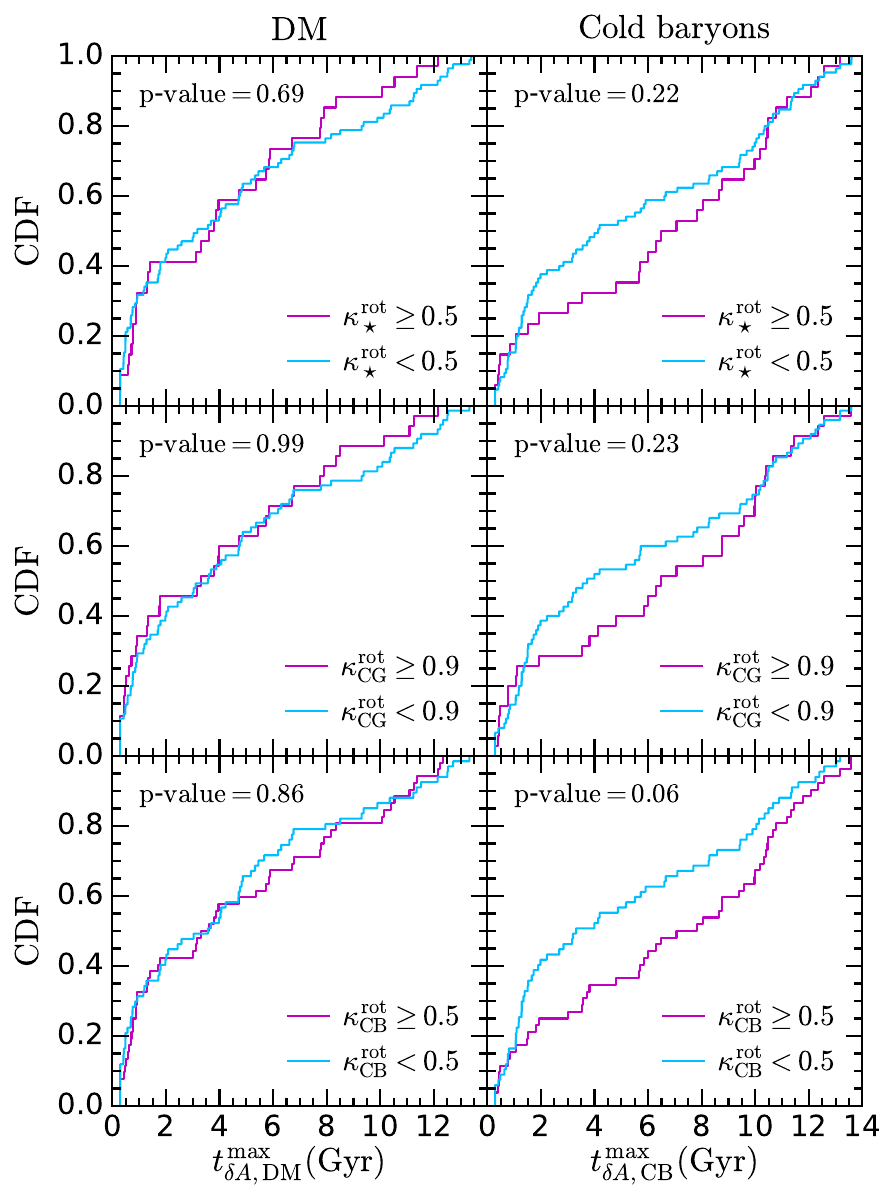} 

\caption{CDFs for  $\tdAmax$ binned by $\kappa^{\rm rot}$ for DM-LVs (left panels) and CB-LVs (right panels). From top to bottom results for $\kappa^{\rm rot}_{\star}$, $\kappa^{\rm rot}_{\rm CG}$, and $\kappa^{\rm rot}_{\rm CB}$ are shown.} 
\label{fig:fot-CDF-KrotBin}
\end{figure}

The effect of the freezing out timescale $\tdAmax$  on the galaxy $\kappa^{\rm rot}$ kinematic parameters is examined in Fig.~\ref{fig:fot-CDF-KrotBin}, where $\tdAmax$ 
CDFs  binned by this  parameter calculated using either stars, cold gas or CB (panels in rows from top to bottom) are shown. 
To account for particle component effects, LVs have been analysed using either their DM (left column panels) or CB (right column panels) particles. 

\paragraph*{Eigen-direction  freezing out timescales.-}

The most statistically significant effect appears in CB-LVs,  when the LV set is split according to the $\kappa^{\rm rot}_{\rm CB}$ kinematic morphological parameter (bottom panel on the right).
Indeed,  highly ordered rotators (magenta lines in Fig.~\ref{fig:fot-CDF-KrotBin}) tend to have been hosted by LVs whose  $\tdAmaxCB$ appear significantly delayed (two-sample KS test with p-value=0.06) relative to CB-LVs where dispersion-dominated galaxies form.  For example,  $40\%$ of CB-LVs with $\tdAmaxCB < 2 \Gyr$ host dispersion-dominated central galaxies, while only  $20\%$ harbour rotationally supported  galaxies. 
This effect is also present when the LV set is binned in 
$\kappa^{\rm rot}_{\star}$ and $\kappa^{\rm rot}_{\rm CG}$ (top and middle panels on the right).  

However, this effect vanishes when the DM-LVs are analysed, thereby implying that the difference between the DM and CB particle behaviour is the responsible for the delay in the $\tdAmax$ distribution  and the  high $\kappa^{\rm rot}_{\rm CB}$ values of  rotation-dominated galaxies. 
 Note that it is the difference between CDFs  for  rotationally supported galaxies (magenta lines; a delay in the skeleton appearance in the case of CB-LVs relative to DM-LVs) what is important.

How can we explain  the aforementioned delay found in the $\tdAmaxCB$ CDF for rotationally supported galaxies relative to kinematic spheroids? This could just be a consequence of the  previously analysed effect of the LV shape.  
We remind the reader that (i) most ordered rotators 
form within highly elliptical CB-LVs, with  $e_{\rm CB}> 0.6$, see Fig.~\ref{fig:epT-CDF-KrotBin}, where we can also see that CB-LVs with $e_{\rm CB}< 0.6$  mostly host kinematic spheroids ($\sim 15 \%$ rotators and $70\%$ kinematic spheroids), and (ii) the emergence of the spines of  highly elliptical CB-LVs is delayed with respect to those of LVs with $e< 0.6$, see Fig.~\ref{fig:CDF-LV-pr}, second row.

\paragraph*{Principal axis  freezing out timescales.-}

Regarding the $\tfmin$ timescale (not shown) the most remarkable result is that its early peak is contributed almost equally by both, rotation and dispersion dominated galaxies. This is, no effect appears on $\kappa^{\rm rot}$. 
 An effect (with p-value=0.21, not very strong from the two-sample KS test)  appears for $\tfmax$ when DM-LVs are binned in  $\kappa^{\rm rot}_{\rm CB}$. 
In this case, no-rotators tend to form within DM-LVs that continue to be for a bit longer under the influence of anisotropic DM mass flows than those associated to highly ordered rotators. 
No other effects have been found.

\subsection{Do the identities of the first or last eigenvector/principal axis to freeze-out affect the host galaxy properties?}
\label{sec:f-oOrderingEffects}

\begin{figure}
\centering
\includegraphics[width=0.65\columnwidth]{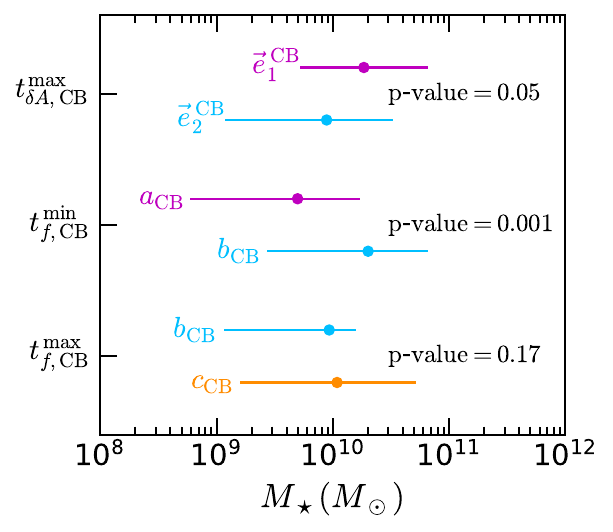} 
    \includegraphics[width=0.65\columnwidth]{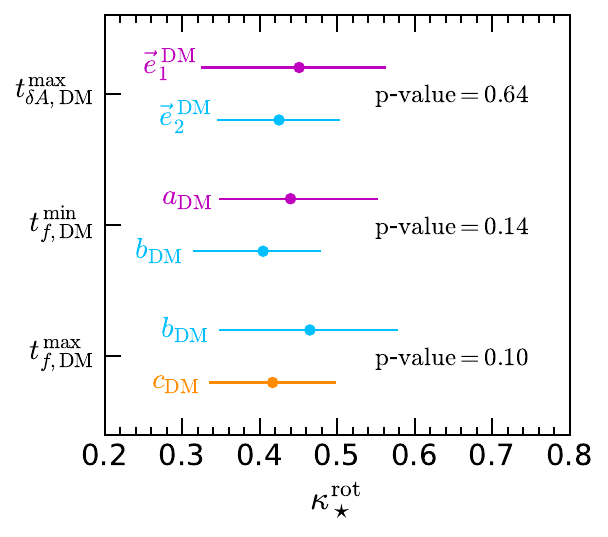}  
\caption{Schematic representation of the CDFs for the stellar mass (upper panel) and the kinematic morphological parameter (bottom panel). In each row, points represent the CDF median values, and the horizontal bars denote the 25\% and 75\% percentiles. Colours stand for the eigen-directions or principal axis. The LV sample has been split according to the last timescale to became frozen ($\tdAmax$ and $\tfmax$, first and third rows respectively), and the first one to freeze out ($\tfmin$, middle row). The p-values of the two-sample KS test are also shown. The freezing out timescales have been calculated using the LV DM particles (bottom panel) or the CB-particles (top panel).}
\label{fig:EsqCDF-MKrot-binned-first-last-frozen}
\end{figure}

 In Section~\ref{src:Ordering-fo-LVs}, we raised the question about the extent to which the ordering of the eigen-directions and principal axes  freezing-out could affect the mass of the LVs.
 Figure~\ref{fig:EsqCDF-MKrot-binned-first-last-frozen} (upper panel) addresses this question, where we see statistically significant stellar mass differences between those galaxies born within the subset of  LVs where the first/last timescale to freeze-out is one or the other (see colours and legends in the plot). 
 
 Possible explanations of these differences come from results of previous sections, and they are as follows:
(i) First row: 
when for a given CB-LV the last eigen-direction to freeze-out is $\vec{e}_1$, then the CB-LV skeleton has emerged at $t_{\rm U} < 2\Gyr$  (see Section~\ref{src:Ordering-fo-LVs}), implying that the CB-LV shape  has a tendency of being  not elliptical (see Fig.~\ref{fig:CDF-LV-pr}, right panel in the second row). 
Therefore, the CB-LV has a higher probability of hosting a massive central galaxy. 
(ii) Second row: when for a given LV the first principal axis to stop changing is $b$, then the  LV is not prolate, and the central galaxy has a higher probability of being massive.
(iii) Third row: if the last principal axis to freeze out is $c$ (orange line), then the timescale $\tfmaxCB$  accumulates at $t_{\rm U} > 11\Gyr$ (see Fig.~\ref{fig:FirstToFrozen-LV} bottom right panel),  so that these CB-LVs have a higher probability of hosting massive central galaxies.

Giving the tight correlation between $\mstar$ and $M_{\rm LV}$ (see Fig.~\ref{fig:MstarMLV}), the same relationships hold between $M_{\rm LV}$  and the principal direction/axis freezing out  timescale ordering.

Some effects on the $\kappa_{\star}^{\rm rot}$ kinematic morphological parameter have also been found in DM-LVs, see bottom panel in  Fig.~\ref{fig:EsqCDF-MKrot-binned-first-last-frozen}. Ordered rotators (i.e., those galaxies showing the highest $\kappa_{\star}^{\rm rot}$ parameter values) appear only in LVs such that either 
the first principal axis to stop changing is $a(t)$ (see second row), or, 
the last mass flows come preferentially from  the $\vec{e}_2^{\rm DM}$ direction, measured through $\tfminDM$ (see third row). Note however that these  effects are weak (as the median $\kappa_{\star}^{\rm rot}$ values are in both cases $< 0.5$), and not that highly significant  using the KS two-sample test, see p-values.

\section{Discussion}
\label{sec:Discussion}

\subsection{How rotational support might have emerged}
\label{sec:RotSupportEmer}


\begin{figure*}
\centering
\includegraphics[width=0.8\textwidth]{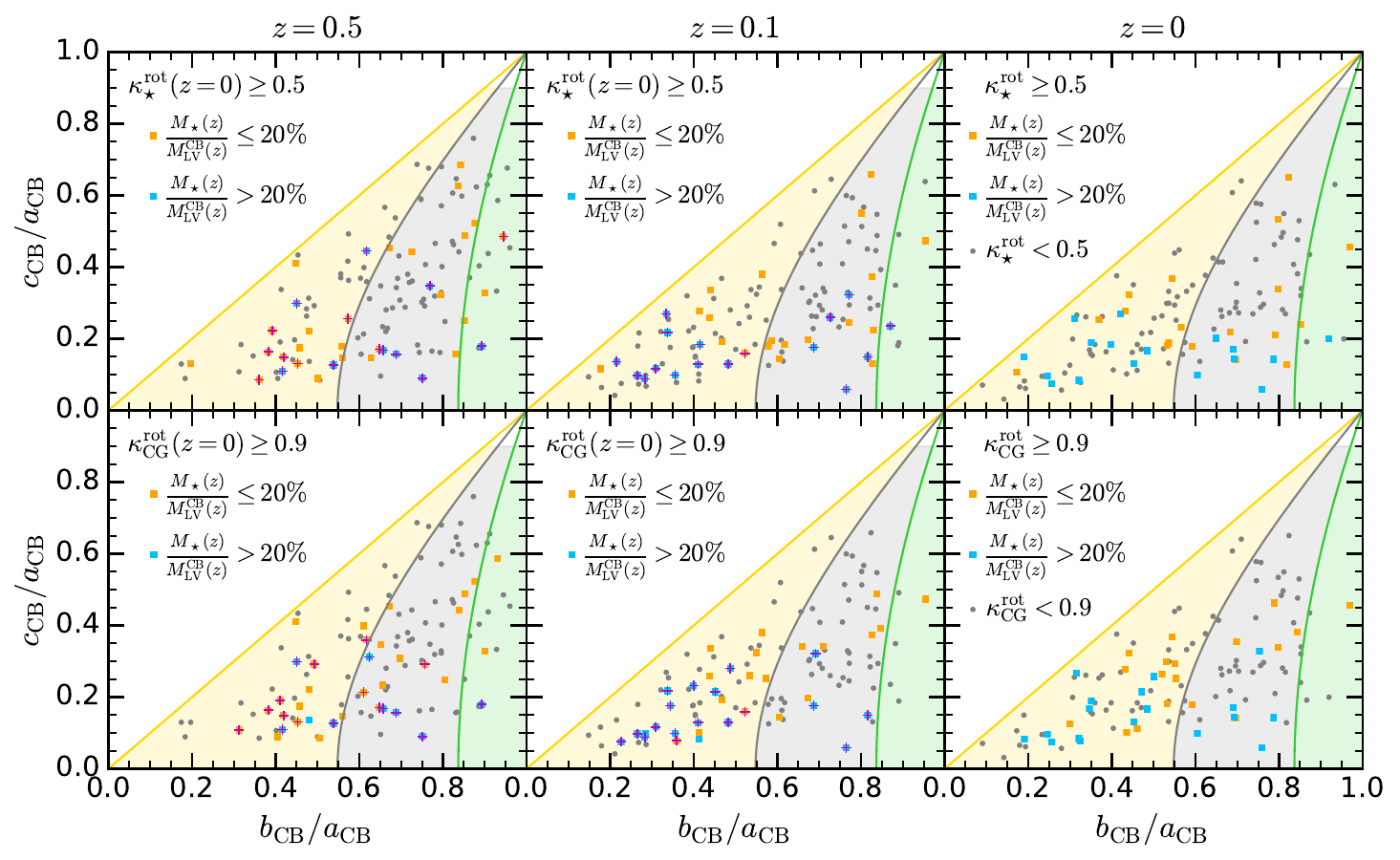}
\caption{CB LVs in the axis-ratio plane binned by the ratio of the central halo stellar mass to the CB mass in each LV for $z=0.5$ (left panel), $z=0.1$ (middle panel) and $z=0$ (right panel).  
Square symbols represent CB-LVs whose central galaxies are supported by rotation (i.e., discy), either when rotation support is defined by $\kappa^{\rm rot}_{\star} > 0.5$ (top panels) or $\kappa^{\rm rot}_{\rm CG} > 0.9$ (bottom panels). 
Grey circles indicate CB-LVs whose central galaxies are dispersion-dominated. 
CB-LVs for which this ratio is $\le20\%$ and that host rotation-dominated galaxies are depicted in orange and those with a ratio $>20\%$ in light blue. 
Magenta symbols: CB-LVs  whose central galaxies are discy at $z=0$ and that have $M_\star/\MLV^{\rm CB}>20\%$ at $z=0$.
}
\label{fig:BuildingRotSupport}
\end{figure*}

Results reported in Section \ref{sec:KrotCentrals} pose the question of how rotation support in discy galaxies is built up.
Figure~\ref{fig:BuildingRotSupport} gives a possible explanation. 
First row panels  indicate that  those central galaxies whose stars are rotationally supported ($\kappa^{\rm rot}_{\star} > 0.5$) have a young stellar population, as at $z=0.5$ the number of systems whose ratio of the central halo stellar mass ($M_\star$) to the CB mass within the corresponding LV ($\MLV^{\rm CB}$) that exceeds the $20\%$ (light blue square symbols) is much lower than at $z=0$ (right panel). This is somewhat expected, as stellar populations become dynamically hotter as time elapses.
The right panel informs us that discy central galaxies with young populations live at $z=0$ within CB-LVs with low  $c_{\rm CB}/a_{\rm CB}$ axis-ratio ($< 0.25$ for all the  cases, and $< 0.20$ except for two cases).  Moreover, except for seven cases, these CB-LVs have $b_{\rm CB}/a_{\rm CB} < 0.5$, that is, they are extremely prolate.

The next question is how this has happened. To answer  this question, in the first and second columns of the first row, redshifts $z=0.5$ and $z=0.1$, respectively,  we plot as magenta crosses  the systems with  $\kappa^{\rm rot}_{\star} > 0.5$ and  $M_\star/\MLV^{\rm CB}>20\%$ at $z=0$ (light blue squares in the right panel). We clearly see that at $z=0.5$ less of these central galaxies were  in   extremely prolate CB-LVs (in comparison to $z=0$), while at $z=0.1$ most of them already are in this kind of LVs, as well as in CB-LVs with low $c_{\rm CB}/a_{\rm CB}$. 
This behaviour strongly suggests that stars have been formed out of cold gas assembled and organized as a thin, flattened   structure with low $c_{\rm CB}/a_{\rm CB}$, a structure that fed on first  a  highly rotation-supported galactic gaseous disc. In addition, most LV shape transformations and star formation occur, in these cases, between $z=0.5$ and $z=0.1$.

To better understand this possible process for stellar rotation support at $z=0$, the same analysis has been repeated for highly ordered rotating cold gas  (i.e., analysing CB-LVs whose central galaxies have  $\kappa^{\rm rot}_{\rm CG} > 0.9$ at $z=0$). Results are given in the second row of Fig.~\ref{fig:BuildingRotSupport}. A comparison of plots on the right panels shows that the central galaxies of 17 CB-LVs are highly ordered rotators both when analysed using stars or CG. In other words, 17  ($\sim 50\%$ of them, see Table \ref{tab:rotators})
of the highly-rotating stellar discs spatially coexist with highly-rotating gaseous discs at $z=0$ in these cases. The same is true at $z=0.5$ and $z=0.1$.  This suggests the following formation sequence: LV  gas organized in a thin, flattened configuration whose kinematic coherent gas fed on a  gaseous galaxy disc in highly ordered rotation from which stars form, giving rise to a stellar disc in ordered rotation.

\subsection{Comparison with other work}
\label{sec:Comparisons}

In this section, we first compare the results of the local environment evolution around galaxy formation sites in EAGLE50, and later, we compare effects of this evolution on mass and rotational support of galaxies.  Due to the specific method used here, only some indirect comparisons are possible in most cases.

A first analysis of the evolution of the density field around galaxy formation sites was carried out by \cite{Robles:2015}. The r-TOI method was applied to a large volume hydrodynamical simulation where  explicit feedback implementation 
was absent, and where a singled out study of DM- and CB-LVs was not carried out. 
Our results  concerning shape parameters evolution agree with theirs,  including  the higher anisotropic configurations  CB-LVs reach as compared to DM- ones in the far from sphericity cases. 
Timescales distributions and their mutual relationships agree as well. The effect of the LV shape on its central galaxy mass is also perfectly consistent.   

When analysing the evolution of the galaxy formation sites at LV scales, we have found that gas does not follow dark matter in EAGLE50. Neither   principal directions and axes of the r-TOI, nor their freezing out timescales. Additionally, as just mentioned,  in the far-from-sphericity regime, the CB LV component has been found to form flattened structures (including highly prolate ones), more anisotropic  than their DM counterparts in the same LV. 

These results are consistent with  analyses of large volume hydrodynamic simulations, such as  that of the TNG300 simulation  \citep[see ][]{Martizzi2019,Artale2022,Walker2024}.
As discussed in Section~\ref{sec:simusobserv},  
 a  flat-shaped, rotationally-organized CGM, fed on by coplanar cold streams,  has been found in different simulations.  This is also in accordance with our findings. 

Our findings also agree with observational results,  mentioned in Section~\ref{sec:simusobserv}, see e.g. \cite{Connor2025}, where a tendency for most of the $z=0$ cool, diffuse IGM being located in voids and sheet-like structures has been pointed out  
 \citep[see also ][]{Tornotti2025b,Tornotti2025}.

 As for other possible  roles played  by wall-like structures made of CB, the so-called Local Sheet \citep{McCall2014} proves the existence of thin flat configurations of baryonic matter in the nearby galaxy distribution, encompassing  the `Council of (massive) Giants', a flat ring-like galaxy configuration with the Local Group at its centre. 
The Council was probably formed on a pre-existing flattened  matter configuration, from which matter was very efficiently accreted. 
Similarly, even if rare, massive galaxy sets spanning flattened structures at $z=0$ have been found in our analysis of EAGLE50,
 and  wall-like CB-configurations are present at early times (see e.g. Fig.~\ref{fig:LVevol-plane} at $z=1$). Our results are also in line with  \citet{Gamez-Marin2024} (and Gámez-Marín, in preparation) physical explanation of satellite kinematic planes, similar to that observed in the MW, as a fossil of a pre-existing early flattened mass configuration.

Another interesting result is that most DM-LVs ($\sim 70\%$) have been found here to evolve into prolate configurations at $z=0$, from which $\sim 46\%$ are extremely prolate, i.e., have prolateness $p > 0.5$, this is  quasi filamentary structures. 
This  preponderance of filamentary structures had already been pointed out  by  \citet{Icke:1972,Icke1984} from a theoretical basis, and found in the first 3D simulations \citep{Centrella:1983,Klypin:1983}.
For a more recent account, see \citet{Cautun:2014}.




Let us now turn to the comparison of our results  to studies of the CW effects on \emph{galaxy} mass and rotation properties.
As previously mentioned, the method usually employed in the literature  involves classifying the CW into its  elements (voids, walls, filaments and knots, in ordered sequence), and then looking  into how mass-related or spin-related properties change according to galaxy location within different CW elements (and to their distance to them).

\citet{Metuki2015}  used the velocity shear tensor
formulation (V-Web method; \citet{Hoffman2012,Libeskind12}),  to classify the CW components. They found that the  more massive a halo (or the galaxy stellar mass) is, 
the more likely it is to be located high in the CW sequence.
 Similar results have been obtained by \citet{Martizzi2020} in their analysis of the TNG100 simulation using a Hessian-matrix-based method to classify the CW  into its component elements. 
Both of these analyses are consistent with our results that point out  CB-LVs evolving into low-$p_{\rm CB}$-shaped structures tend to host 
massive galaxies (and haloes), i.e. with large $\mstar$.
Also, with those CB-LVs acquiring extremely prolate shapes are prone to host low $\mstar$ galaxies (and haloes).
In addition, we have found similar  results  when analysing DM-LVs, albeit the effects are less strong than those found in CB-LVs.




Regarding rotation properties of galaxies, an important result is that rotational support cannot be explained as driven by galaxy (or halo) mass, in consistency with \citet{Correa:2017} and \citet{Celiz:2025} results for the EAGLE100 and TNG50 simulations, respectively, within the stellar mass range analysed here (see Fig. \ref{fig:Krot-vs-mass}). 
 Interestingly, a remarkable effect of the LV shape on central galaxy rotational support has been identified in CB-LVs through the ellipticity parameter and  the $(c/a)_{\rm CB}$ axis-ratio (whose values are closely related). In DM-LVs the shape effect has been detected using the triaxiality and prolateness parameters. This relationship is compatible with \citet{Sales:2012} proposal. 


\section{Summary and Conclusions}
\label{sec:SumConclu}

In this paper, we investigate the evolution of the environment of galaxy formation sites. As a first application of our results, we analyse the extent to which the characteristics of this evolution determine two basic properties of the galaxies at low redshift, namely  mass and rotation support.

\subsection{Summary of methods}
\label{sec:SummMethods}

To this aim, we used the EAGLE reference simulation of 50 Mpc box length.  
For each central galaxy in the simulation, particles within a spherical region of comoving radius ten times the virial radius at $z=0$, with centre at the proto-galaxy formation site,  was singled out at high redshift ($z=15$). Then,  the evolution of these particles that constitute  a Lagrangian Volume (LV) was traced down to $z=0$,  and the deformation of each LV  quantified using its reduced  inertia tensor (r-TOI).
The evolution of the environment around forming galaxies   has been described in terms of two groups of parameters: (i) those related to the shape deformation of the initially spherical LVs, encoded in  the r-TOI principal axes ($a(t),b(t)$ and $c(t)$)) at a given cosmic age, $t$, and in their derived shape parameters: triaxiality, $T(t)$, ellipticity, $e(t)$, and prolateness $p(t)$), as well as in the r-TOI principal directions, and  
(ii)  the timescales of two important events for galaxy formation relative to its environment, namely: the age of the Universe $t_{\rm U}$  when the local spine emerges, $\tdAmax$  (i.e., when the mass flow directions towards the formation sites become fixed in time within  $10\%$), and the age of the Universe  when anisotropic mass rearrangements around the forming galaxy vanish at the LV scale, $\tfmax$. 
It is also interesting to study the timescales for the freezing out of the first eigen-direction, $\tdAmin$, and the first principal axis, $\tfmin$.

We have analysed the deformations of these LVs using their corresponding r-TOIs, calculated by sampling each LV with its dark matter (DM) and cold baryons (CB) constituent particles (referred to as DM-LVs and CB-LVs).
The evolution of the DM-LVs is determined by the effect of  dynamical forces, as in the Zel’dovich  Approximation (ZA) and its extension to the Adhesion Model (AM). For the CB component, hydrodynamic processes as well as the effects of energy injection by discrete sources (supernovae, black holes), and the subsequent thermodynamic processes, acting at sub-resolution scales,  come  into play as well. 

It is worth reminding the reader that most methods to analyse the cosmic web  evolution in the literature use functions defined at each mass element (or point), and  a smoothing procedure has to be applied in order to make  calculations. Conversely, the reduced  inertia tensor analysis provides a global description of the local environment  at the LV scale (i.e., adapted to the particular object under consideration). This is simpler than the usual methods since, in practice, it requires only four basic time-dependent shape functions, one of which freezes-out very soon. 
Our method identifies a direction of maximum compression in most cases.
In practice, this direction is the same as that returned by other  methods, such as the velocity shear tensor analysis. While the deformation or the velocity shear tensors are a more appropriate scheme when trying to detect and classify cosmic web structures  \citep[see e.g. ][]{Forero-Romero:2009,Hoffman2012,Cautun:2014,Libeskind18}, both formalisms, ours included, allow to identify the main directions of mass flows.


\subsection{Summary of main results and conclusions}

These are the main results of this work regarding the evolution of the environment around galaxy formation sites:

\begin{enumerate} 

\item Environments around galaxy formation sites generally evolve into highly anisotropic configurations. 67\% of the  initially spherical DM-LVs at $z=15$ become highly prolate at $z=0$, and only 4\% of them are oblate.
Frequently a DM-LV first rapidly flattens, then it becomes  
triaxial and finally more prolate. 
In the far-from-sphericity regime, the evolution of CB-LVs usually results in even more anisotropic configurations at low redshift than that of their DM counterparts.
\item 
Both the $z=0$ halo and stellar mass of a given central galaxy depend on  the shape of the LV that harbours them. 
CB-LVs acquiring extremely prolate shapes  show a clear tendency to host low $\mvirzlow$  haloes. Those evolving into low-$p$ CB-LVs  host preferentially massive haloes at $z=0$.
Similar  results are found when analysing DM-LVs,  slightly less strong though.
The effect of the LV shape  measured via the triaxiality $T$ parameter is outstanding in that low $\mvirzlow$ haloes most likely form in prolate-like  LVs ($T > 0.7$), while massive haloes are not clearly prone to appear in a prolate or a triaxial LV. 
\item LV shape evolution takes place at two different paces:  very fast at early times, then it slows down until it freezes out. In general, eigen-directions freeze out 
before  principal axes ($\tfmax$ peaks at $\sim 11 \Gyr$).
\item Around $100\%$ of the DM- or CB-LVs have one of their principal directions fixed in the first $2\Gyr$ of $t_{\rm U}$. In most cases, the  eigen-direction of the major axis is fixed first, irrespective of the analysed component, DM or CB.
Concerning the setting in of their spine, at $z=0.5$ $\sim 40\%$ of the LVs have not reached  their final directional configuration yet. 

\item LVs hosting massive  haloes (with large $\mvirzlow$)
tend to have their spine fixed earlier on than those containing less massive haloes. This is a consequence of the ZA and AM. 

\item The first principal axis to freeze-out is in general $b(t)$, in $\sim 30\%$ of the LVs this occurs within the $\sim 2 \Gyr$. 
 Note that LVs that fix their  intermediate principal axis,  $b(t)$, very early
 cannot acquire a prolate shape by evolution.  They have $T < 0.7$,   
 and they host a few low-mass  haloes at $z=0$.
Either for DM- or CB-analysed LVs, the last principal axis to stop changing is, in most cases, the minor axis $c(t)$  (associated to the most compressive  mass flows),
\item The end of DM-mass flows  
occurs late on average (at $t_{\rm U}\sim 10.5 \Gyr$, and even slightly later for CB-LVs). For low-mass haloes, this endpoint is spread in the $t_{\rm U}\sim  [6 - 13] \Gyr$ interval.

\item The shapes of both, DM-LVs  and CB-LVs, at $z=0$ are strongly related to their freezing out timescales. Statistically, highly anisotropic CB-LVs have both their spine setting in and the end of their anisotropic mass reconfigurations notably retarded with respect to less anisotropic configurations, as well as with respect to those of DM-LVs.
Conversely, at a high statistical significance,  the spine of highly anisotropic DM-LVs emerges  
before than that of less anisotropic DM-LVs. Thereby, anisotropic LVs do not fix their CB spines until several Gyr after their DM spines have emerged.

 \end{enumerate}

 We now turn to summarize the effects of environment evolution on galaxy mass and galaxy rotation support:

 \begin{enumerate}
 
 \item Via the  stellar-to-halo mass relation, SHMR, the relations described above between $\mvirzlow$ and $\tdAmax$,  and between $\mvirzlow$ and $\tfmax$, are translated into relationships between $\mstar$ and $\tdAmax$, and between $\mstar$ and $\tfmax$, respectively. The same can be said of  the relations between  $\mstar$ and  LV shape parameters.

\item   As for rotational support acquisition, at $z=0$ and for either DM- or CB-LVs, highly ordered rotators (i.e., with   kinematic morphological parameter $\kappa^{\rm rot} > 0.5$ for stars and CB, and $> 0.9$ for cold gas) appear in highly elliptical LVs 
(those with 
$(c/a)$ distributions with median values $\sim 0.2$, and narrow dispersions). Conversely, the distribution for dispersion-dominated central galaxies has a wider dispersion  and peaks at higher $(c/a)$ axis-ratios.
These median $(c/a)$ axis ratios decrease from $z=1$ onwards.

\item 
While rotation-dominated galaxies tend to have been hosted by LVs whose $\tdAmax$ timescales in cold baryons are large,  dispersion-dominated galaxies tend to form within LVs whose $\tdAmax$ in CB is shorter.
Conversely, this effect is not present in  DM-LVs, thereby implying that it  is the  CB particle behaviour that causes the delay in the $\tdAmax$ timescale  and the  high $\kappa^{\rm rot}$ values.

\item At $z=0$, most central galaxies with ordered stellar rotation present star formation later than $z=0.5$. Half of them also possess highly rotation-supported gaseous discs (with $\kappa^{\rm rot} > 0.9$) from at least $z=1$.

\end{enumerate}

From this summary of results, we extract and highlight the most outstanding conclusions of this work:
\begin{enumerate}
\item  CBs do not follow the DM evolution around galaxy formation sites, but DM guides the CB cosmic web deployment. While cold gas structures arising from dissipation  are more anisotropic than their DM frames (i.e., filament spines; walls), or more concentrated (galaxies versus haloes), gaseous configurations heated by gravitational or discrete energy injection processes overflow their DM guides.

\item LV shapes affect central galaxy masses. While low-mass central galaxies (or haloes) tend to form in CB-LVs that evolved into highly prolate-shapes, 
massive galaxies  more likely form in less anisotropic CB-LVs. 

\item Wall-like structures of CB appear early in some cases, in consistency with incipient observational results at these redshifts. Flattened current galaxy configurations could be a fossil of these early structures, whose role in rotational support acquisition by central galaxies might have been very relevant.

\item The LV freezing out timescales and central galaxy masses show mutual relations. While  LVs with larger $\tfmax$ timescale are more likely to have fed on massive galaxies, the LV spine setting in DM-LVs tends to be retarded in less massive ones.

\item  Rotational support and LV shape evolution have been found to be related. Indeed,  highly ordered rotators tend to form in low $(c/a)$ CB-LVs, including highly prolate ones, while dispersion-dominated central galaxies more likely form in close to spherically shaped DM-LVs.

\item  The  highest statistically significant  difference between the DM and CB behaviour found here is the delay of the $\tfmaxCB$ distribution for massive objects, relative to both, less massive objects and to the $\tfmaxDM$ distributions for massive objects.

\end{enumerate}

Finally, our results show that the reduced inertia formalism is an appropriate tool to analyse the evolution of the local environment around galaxy formation sites (i.e., its shape structuring and timescales). 
To the best of our knowledge, this is the first time that the evolution of the environments of galaxy formation sites is investigated considering  feedback effects on the baryonic component. 

An extension of the work presented here is needed to dig deeper into a more complete catalogue of galaxy properties, a task requiring higher resolution and higher volume hydrodynamical simulations.

 \section*{Acknowledgements}
SR  was partially supported by the UK STFC grant ST/T000759/1 and by the Fermi National Accelerator Laboratory (Fermilab), a U.S.
Department of Energy, Office of Science, HEP User Facility. 
This work was performed in part at Aspen Center for Physics, which is supported by National Science Foundation grant PHY-2210452. 
SR thanks the Theoretical Physics Department at Universidad Auton\'oma de Madrid, Spain, for its kind hospitality while this work was being conducted. 
SR also  acknowledges CERN TH Department for its hospitality while this research was being carried out. 
%
SEP acknowledges support from MinCyT (Argentina) through BID PICT 202000582.
We thank the Ministerio de Ciencia e Innovación (Spain)  for financial support under Project grant PID2021-122603NB-C21,
as well as Project  PID2024-156100NB-C21  financed by MICIU/AEI
/10.13039/501100011033 / FEDER, EU.
 This work has received funding from the European Union's HORIZON-MSCA-2021-SE-01 Research and Innovation programme under the Marie Sklodowska-Curie grant agreement number 101086388 - Project acronym: LACEGAL.
This research was undertaken using the Computational Research, Engineering and Technology Environment (CREATE) at \citet{CREATEHPC}, and Wilson Cluster at Fermilab.

\section*{Data Availability}
The data underlying this article will be shared on reasonable request to the corresponding author.
 



\bibliographystyle{mnras}
\bibliography{references} 


%

\appendix

\section{Complementary Figures}
\label{app:appendixA}

\begin{figure}
\centering
\includegraphics[width=\columnwidth]{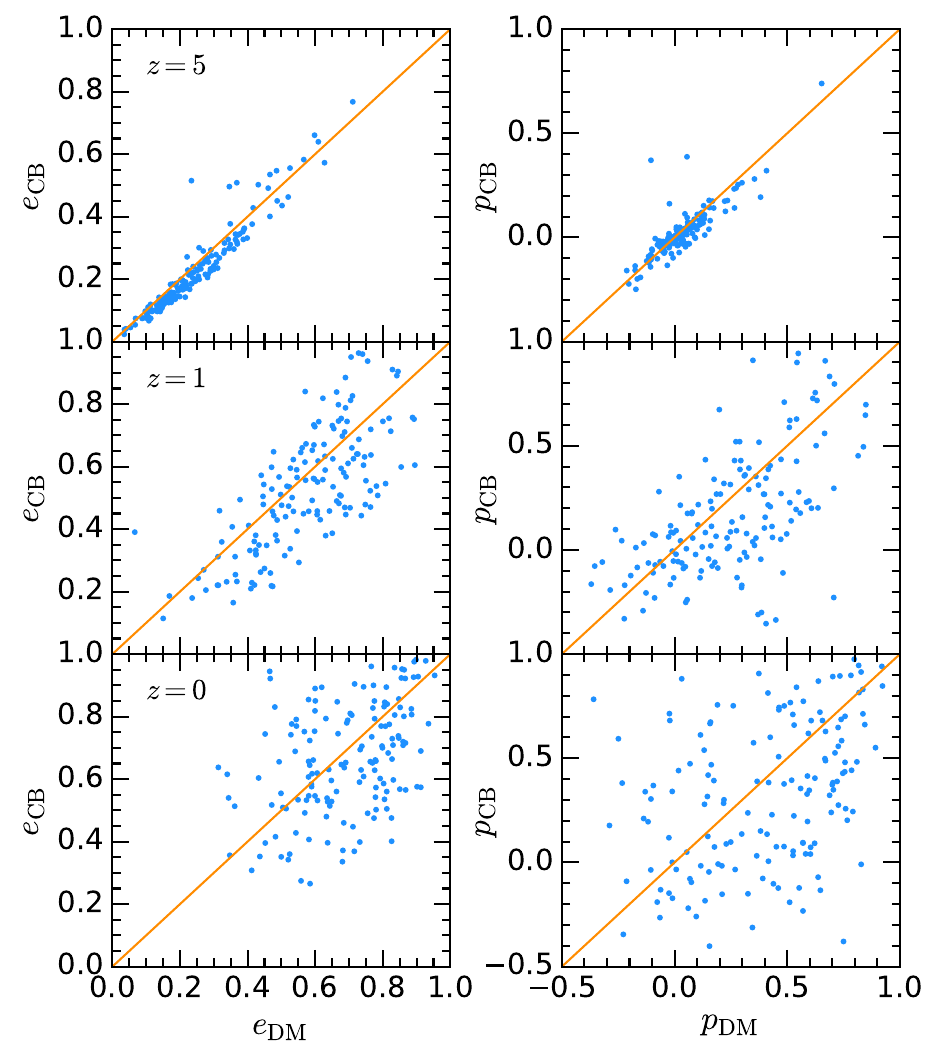}
\caption{Evolution of the ellipticity (left panels) and prolateness  (right panels) for the cold baryon reduced inertia tensor with respect to that of the DM component. }
\label{fig:epT-DM-CB}
\end{figure}

In Fig.~\ref{fig:epT-DM-CB}, the $e$ and $p$ shape parameters for the CB reduced inertia tensor are compared against those of their corresponding DM-LV counterparts. While we see that few structures reach $e_{\rm DM} > 0.8$ or 
$p_{\rm DM} > 0.8$ at $z=0$, their CB counterpart do it to a larger extent. Therefore, in the case of far-from-sphericity structures, the CB-LVs reach more anisotropic configurations than their DM-LV counterparts. 

\begin{figure}
    \centering  
   \includegraphics[width=0.9\columnwidth]{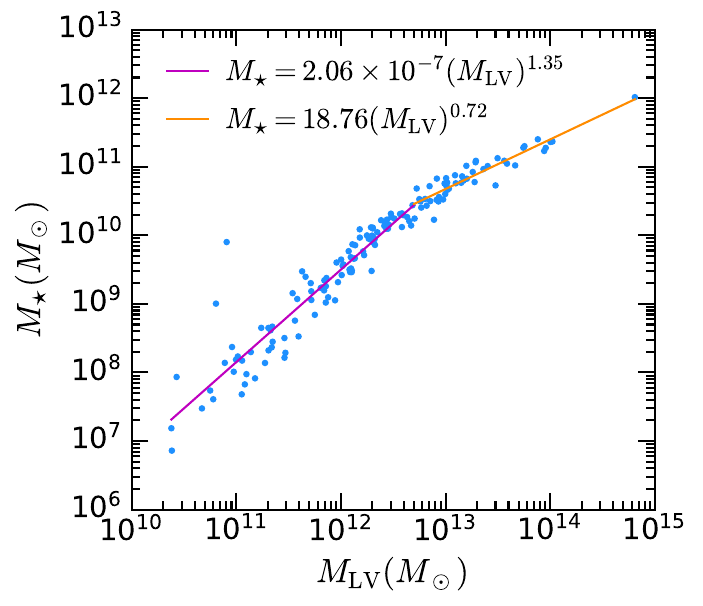}   
    \caption{Galaxy stellar mass versus LV mass.}         
    \label{fig:MstarMLV}
\end{figure}

A tight correlation between $\mstar$ and $M_{\rm LV}$, and, consequently, between $\mstar$ and $\mvirzlow$   is shown in Fig.~\ref{fig:MstarMLV}. This is the so-called stellar-to-halo mass relation \citep[see e.g, ][]{Wechsler:2018}. This is used in this paper to translate relations involving the halo mass, $\mvirzlow$, to those involving the galaxy stellar mass, $\mstar$. 


\begin{figure}
\centering
\includegraphics[width=0.8\columnwidth]{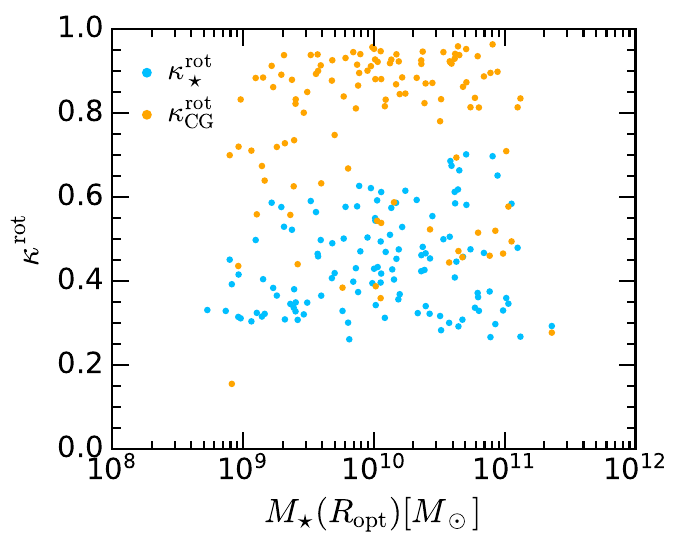}
\includegraphics[width=0.8\columnwidth]{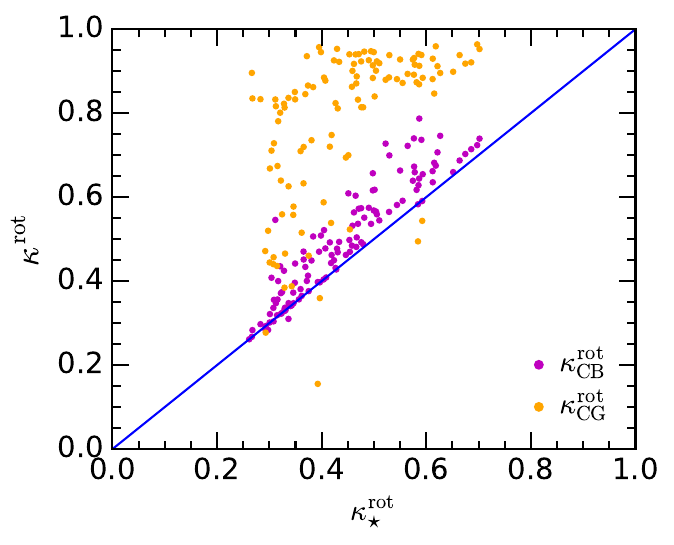}
\includegraphics[width=0.8\columnwidth]{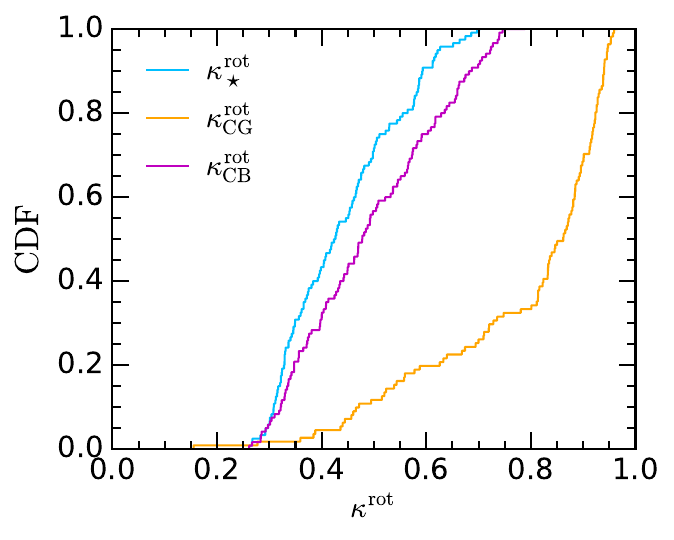}
\caption{Top: $\kappa^{\rm rot}_\star$ and  $\kappa^{\rm rot}_\text{CG}$ as a function of the stellar mass with the optical radius $M_\star(\ropt)$. Middle: $\kappa^{\rm rot}_\text{CG}$ and $\kappa^{\rm rot}_\text{CB}$ versus $\kappa^{\rm rot}_\star$. Bottom: CDF for the different $\kappa^{\rm rot}$ parameters.}
\label{fig:Krot-vs-mass}
\end{figure}

The $\kappa^{\rm rot}$  parameter has been calculated for the galaxies in our sample,
  out of either stellar, cold gas or cold baryon  particles ($\kappa^{\rm rot}_{\star}$,
$\kappa^{\rm rot}_{\rm CG}$ and $\kappa^{\rm rot}_{\rm CB}$, respectively). 
Correlations among them  
can be found in Fig.~\ref{fig:Krot-vs-mass}. As expected, $\kappa^{\rm rot}_{\rm CB}$  is always higher than $\kappa^{\rm rot}_{\star}$ for stellar discs 
(see bottom panel), since cold gas contributes highly ordered motion (i.e., roughly circular and in-plane orbits)  to kinematic discs. Circular and in-plane orbits are those endowed with the minimum mechanical energy in an axial potential \citep[see][]{BinneyTremaine08}. In addition, 
 stellar discs host highly ordered gaseous 
discs, as seen in the middle panel.
We also note in the same plot that  kinematic stellar spheroids (i.e., galaxies with $\kappa^{\rm rot}_{\star} < 0.5$) harbour gaseous structures with
a wide range of $\kappa^{\rm rot}_{\rm CG}$ values, being either gaseous discs or not.

In the top panel of Fig.~\ref{fig:Krot-vs-mass}, we plot $\kappa^{\rm rot}_{\star}$ and  $\kappa^{\rm rot}_{\rm CG}$ 
versus galaxy stellar masses within one optical radius $\ropt$. 
It is worth noting that the $\kappa^{\rm rot}_{\star}$ versus $\mstar$ relation is consistent with those of \citet{Correa:2017} and \citet{Celiz:2025} for the EAGLE100 and  TNG50 simulation, respectively, within the  galaxy stellar mass range analysed here. 
  No correlation appears between the $\kappa^{\rm rot}_{\star}$   parameter and  the stellar mass of the resulting galaxy (light blue points), except 
for those galaxies with $M_\star(R_{\rm opt}) < 10^9 M_{\odot}$ which have $\kappa^{\rm rot}_{\star} \sim 0.3 - 0.4$, i.e., they are not kinematic rotators.   
Moreover, there are only a few rotationally supported cold gas structures for these relatively low stellar mass central galaxies in the EAGLE50 reference simulation,
see the $\kappa^{\rm rot}_{\rm CG}$  versus   $M_{\star}(R_{\rm opt})$ (orange circles). Conversely, the same plot indicates that  in many cases galaxies  with 
$M_\star(R_{\rm opt}) > 2 \times 10^9 M_{\odot}$ have $\kappa^{\rm rot}_{\rm CG} \sim 0.9$ , i.e., they host  highly rotationally supported gaseous 
discs.

As mentioned above, correlations of 
 $\kappa^{\rm rot}$ with stellar galaxy mass are weak. This is because   the former  largely depends on the extent to which  the
angular momentum of baryons being accreted over time, and that feed on the proto-galaxy, is not only high in magnitude, but also aligns coherently with each other 
so that their vectorial sum does not add up to a low value at accretion time. 


Finally, a list of the terms used in this work is given in Table~\ref{tab:Glossary}. 

\begin{table*}
\centering
\caption{Terms used in this work.}
\label{tab:glossary}
\begin{tabularx}{\textwidth}{lX @{}}
\hline
  \textbf{Term} & \textbf{Definition} \\
\hline
LV & Lagrangian Volume, see  Section \ref{sec:maths}.\\
r-TOI         & Reduced Tensor of Inertia defined in Eq. \ref{eq:redineten}, considering all the particles within the LV.
\\
DM, CG, $\star$, CB    & Type of component: dark matter, cold gas, stellar, cold baryon (made of CG plus stars) respectively.\\
comp-LV & LV where only its comp. (DM, CB) particle components are considered. \\
 r-TOI-DM, r-TOI-CB  & Reduced Tensor of Inertia calculated with either DM or CB LV particles, respectively. \\
 $ \hat{e}_i^{\rm comp}(t)$, $i=1,2,3$ &  The $i$-th principal direction of the r-TOI (calculated using the comp-type particles) at Universe age $t$. \\
$a(t)_{\rm comp}$ & The major principal axis of the r-TOI  for the comp-LV at Universe age $t$.\\
$b(t)_{\rm comp}$ & The intermediate principal axis of the r-TOI  for the comp-LV at Universe age $t$.\\
$c(t)_{\rm comp}$ & The minor principal axis of the r-TOI  for the comp-LV at Universe age $t$.\\

$T_{\rm comp}$ &  Triaxiality parameter, see Eq. \ref{eq:Tdef}, for the comp-LV.\\

$e_{\rm comp}$, $p_{\rm comp}$ &  Ellipticity and prolateness parameters, see Eq. \ref{eq:ShapePDef}, for the comp-LV.\\

$\tdAmin$ & Universe age when just 1 eigen-direction has frozen-out (within a 10\%).\\
$\tdAmax$ & Universe age when the 3 eigen-directions have frozen-out (within a 10\%).\\
$\tfmin$ & Universe age when the first principal axis, either $a, b$ or $c$,  freezes-out (within a 10\%).\\
 $\tfmax$ & Universe age when the 3 principal axes, $a, b$ and $c$,  have frozen-out (within a 10\%).\\
$t_{\delta A, \rm DM}^{\rm min}$, $t_{\delta A, \rm CB}^{\rm min}$ & $\tdAmin$  calculated with either DM or CB particles,  respectively. \\ 
$t_{\delta A, \rm DM}^{\rm max}$, $t_{\delta A, \rm CB}^{\rm max}$  & $\tdAmax$ calculated with either DM or CB particles, respectively.\\
$t_{f, \rm DM}^{\rm min}$, $t_{f, \rm CB}^{\rm min}$  & $\tfmin$ calculated with either DM or CB particles, respectively.\\
$t_{f, \rm DM}^{\rm max}$, $t_{f, \rm CB}^{\rm max}$  & $\tfmax$ calculated with either DM or CB particles, respectively.\\
$\kappa^{\rm rot}$  & Central galaxy mechanical energy fraction in rotational motion,  defined in Eq. \ref{eq:KrotDef} .\\
$\kappa^{\rm rot}_{\star}$, $\kappa^{\rm rot}_{\rm CG}$, $\kappa^{\rm rot}_{\rm CB}$ & $\kappa^{\rm rot}$  calculated using either stars, cold gas, or cold baryons, respectively.\\
\hline  
\end{tabularx}
\label{tab:Glossary}
\end{table*}



\bsp	
\label{lastpage}
\end{document}